\documentclass[11pt]{article}
\usepackage{amsfonts}
\usepackage{amssymb}
\usepackage{amsmath}
\usepackage{amsthm}
\usepackage{epsfig}

\newlength{\bredde}
\def\slash#1{\settowidth{\bredde}{$#1$}\ifmmode\,\raisebox{.15ex}{/}
\hspace*{-\bredde} #1\else$\,\raisebox{.15ex}{/}\hspace*{-\bredde} #1$\fi}
\newcommand{\be}{\begin{equation}}
\newcommand{\ee}{\end{equation}}
\newcommand{\bea}{\begin{eqnarray}}
\newcommand{\eea}{\end{eqnarray}}
\newcommand{\nn}{\nonumber}
\newcommand{\eins}{\leavevmode\hbox{\small1\kern-3.8pt\normalsize1}}
\newcommand{\e}{\mbox{e}}

\newcommand{\sect}[1]{\setcounter{equation}{0}\section{#1}}

\def\Tr{{\mbox{Tr}}}

\def\Zd{{\cal Z}_{\delta,\nu}}
\def\Zdn{{\cal Z}^{\cal N}_{\delta,\nu}}
\def\Z{{\cal Z}_{\nu}}
\def\Zn{{\cal Z}^{\cal N}_{\nu}}
\def\G{{\cal G}}

\begin{document}
\topmargin -1.4cm
\oddsidemargin -0.8cm
\evensidemargin -0.8cm
\title{\large{{\bf
Products of random matrices from fixed trace\\ and induced Ginibre ensembles}}}

\vspace{1.5cm}

\author{~\\{\sc Gernot Akemann} and {\sc Milan Cikovic}
\\~\\
Faculty of Physics,
Bielefeld University,\\
P.O. Box 100131,
D-33501 Bielefeld, Germany
}
\maketitle
\vfill

\begin{abstract}

We investigate the microcanonical version of the complex induced Ginibre ensemble, by introducing a fixed trace constraint for its second moment. 
Like for the canonical Ginibre ensemble, its complex eigenvalues can be interpreted as a two-dimensional Coulomb gas, which are now subject to a constraint and a modified, collective confining potential. Despite the lack of determinantal structure in this fixed trace ensemble, we compute all its density correlation functions at finite matrix size and compare to a fixed trace ensemble of normal matrices, representing a different Coulomb gas. Our main tool of investigation is the Laplace transform, that maps back the fixed trace to the induced Ginibre ensemble.
Products of random matrices have been used to study the Lyapunov and stability exponents for chaotic dynamical systems, where the latter are based on the complex eigenvalues of the product matrix. Because little is known about the universality of the eigenvalue distribution of such product matrices, we then study the product of $m$  induced Ginibre matrices with a fixed trace constraint - which are clearly non-Gaussian - and $M-m$ such Ginibre matrices without constraint. 
Using an $m$-fold inverse Laplace transform, we obtain a concise result for the spectral density of such a mixed product matrix at finite matrix size, for arbitrary 
fixed $m$ and $M$.  Very recently local and global universality was proven by the authors and their coworker for a more general, single elliptic fixed trace ensemble in the bulk of the spectrum. Here, we argue that the spectral density of mixed products is in the same universality class as the product of $M$ independent induced Ginibre ensembles.

\end{abstract}

\vfill

\thispagestyle{empty}
\newpage

\renewcommand{\thefootnote}{\arabic{footnote}}
\setcounter{footnote}{0}

\sect{Introduction}\label{intro}

Both ideas of providing random matrices with a fixed trace constraint, and of multiplying random matrices go back to the early days of random matrix theory (RMT), see Rosenzweig \cite{Rose} and Mehta's classical book \cite{Mehta:2004} in its 1st edition from 1967, and the work of Furstenberg and Kesten \cite{FK}, respectively. 
While fixed trace ensembles were introduced from a statistical physics point of view, yielding a microcanonical version of the classical canonical Gaussian ensembles, products of random matrices were studied as a toy model for dynamical systems with chaotic dynamics, by looking at the Lyapunov or stability exponents of the product matrix. 
The goal of this paper is to bring these topics together in the context of non-Hermitian RMT. Apart from possible applications our main aim is  to find exact and closed form expressions for the complex eigenvalue correlation functions of a single fixed trace ensemble and $M$ products thereof, at finite $M$ and finite matrix size $N$.
Such expressions can then be used to analyse questions of universality in various large-$N$ limits that are relevant for the corresponding two-dimensional Coulomb gas or dynamical systems picture.

Fixed trace ensembles (FTE) are defined by replacing the Gaussian distribution for all independent matrix elements, $P(G)\sim\exp[-t\Tr GG^\dag]$, by a constraint on the second moment enforced by a delta-function, $P_\delta(G)\sim\delta(s-\Tr GG^\dag)$, with $s,t>0$. 
While for standard FTE usually $G$ is Hermitian,  we will chose it to be a complex non-Hermitian rectangular $N\times (N+\nu)$ matrix here and study its complex eigenvalues.
Let us, first, give a very brief account on FTE and their possible applications 
and, second,  on complex eigenvalue distributions of products of random matrices, in order to put our work in a more general context.

FTE were initially introduced by Rosenzweig \cite{Rose} (as cited in \cite[Chapter 27]{Mehta:2004}) for Gaussian Hermitian random matrices, and we will focus on matrices with complex elements only.
The relation between the Gaussian and FTE is in analogy to that between the canonical and micro-canonical ensemble in statistical mechanics. The distinct feature of FTE is that they are compactly supported already at finite-$N$, their corresponding point process of  eigenvalues is non-determinantal, and in particular the distribution of matrix elements is non-Gaussian. 
Like their Gaussian partners the eigenvalues of the FTE can be interpreted as a gas subject to the two-dimensional Coulomb interaction. Due to the Hermiticity of the matrix the eigenvalues are then restricted to the real line. The confining Gaussian potential of the canonical ensemble is then replaced by a delta-function constraint. A further version of so-called restricted trace ensembles exists, cf. \cite[Chapter 27]{Mehta:2004}, where the constraint is given by a Heaviside-Theta function.
When moving to complex non-Hermitian matrices here, and studying their complex eigenvalues, we will thus be able to analyse a true two-dimensional Coulomb gas at a specific inverse temperature $\beta=2$. Such Coulomb gases at general temperatures have been an intense object of study recently, and we refer to \cite{Serfaty} for a recent review. 

There is one subtlety however, when generating two-dimensional Coulomb gases from random matrices. While normal matrices with complex \cite{Ginibre} or quaternion entries \cite{Hastings} indeed correspond to the same Coulomb gas at different inverse temperatures $\beta=2$ and 4, respectively, Gaussian Ginibre matrices with real, complex or quaternion entries lead to {\it different} Coulomb gases, and we refer to \cite{PeterCoulomb} for the interpretation of these different classes of particle systems. Only for complex entries Ginibre and normal matrices share the same complex eigenvalues statistics\footnote{Note however, that their corresponding eigenvector dynamics is different, with only the diffusion of Ginibre matrices coupling to the overlap, cf. \cite{JPB}.}. Is this a coincidence of the Gaussian case? As we will see in this work, after imposing a fixed trace constraint the joint densities of eigenvalues of normal and of Ginibre matrices will correspond to two different two-dimensional Coulomb gases, one with a delta-function constraint and one with a collective potential and Heaviside-constraint, respectively.

A different class of FTE, where $G$ is also complex non-Hermitian, has become popular more recently,  due to their relation to the  
so-called Hilbert-Schmidt eigenvalues of the density operator in bipartite quantum systems, see \cite{Satya} for a review and references. Let us briefly describe the standard setup where usually the singular value statistics of $G$ (or equivalently the eigenvalues of $GG^\dag$) is investigated, and under which circumstances it can become interesting to consider the complex eigenvalues of $G$ instead. 
A bipartite quantum system is given by a Hamiltonian of the form ${\cal H}={\cal H}^A\otimes {\cal H}^B$, where typically one of the systems, say $A$ of dimension $N$,  is viewed as a subsystem, and the other as the environment, that is then $B$ of dimension $M$, where typically $M\geq N$. Both Hamiltonians are characterised by a basis of right (R) and left (L) eigenvectors $\{|R_i^A\rangle\}$ and $\{\langle L_i^A|\}$
for system $A$, and 
$\{|R_\alpha^B\rangle\}$ $\{\langle L_\alpha^B|\}$ for system $B$, which we will normalise for each system, $\langle L_i^A|R_j^A\rangle=\delta_{i,j}$ and $\langle L_\alpha^B|R_\beta^B\rangle=\delta_{\alpha,\beta}$. When both Hamiltonians are Hermitian, left and right eigenvectors agree, $|R_i^A\rangle^\dag=\langle  L_i^A|$, and likewise for $B$. However, if we consider non-Hermitian Hamiltonians e.g. for open systems, the two sets have to be distinguished and a nontrivial overlap amongst left eigenvectors (and amongst right eigenvectors) exists, see e.g. \cite{Mehlig} which has created some renewed interest recently, cf. \cite{YVF,BD,NT}.
In any case a quantum state can be decomposed as $|\varphi\rangle=\sum_{i,\alpha}G_{i,\alpha}|R_i^A\rangle\otimes |R_\alpha^B\rangle$, with coefficients given by a complex $N\times M$ matrix $G$. The density matrix of this state is then given by 
\begin{equation}
\rho = \sum_{i,\alpha}\sum_{j,\beta}G_{i,\alpha}G_{j,\beta}^* |R_i^A\rangle\langle L_j^A|\otimes |R_\alpha^B\rangle\langle L_\beta^B|\ .
\end{equation}
Defining the partial trace by $\rho_A=\mbox{Tr}_B[\rho]=\sum_\alpha\langle L_\alpha^B|\rho   |R_\alpha^B\rangle$, and likewise for $\rho_B$, this density matrix is normalised, 
$\mbox{Tr}_A\mbox{Tr}_B[\rho]=1$. The leads to the constraint 
\begin{equation}
\mbox{Tr}GG^\dag=1\ .
\label{HS}
\end{equation} 
For Hermitian Hamiltonians one usually changes basis to the singular values of $G$ and expresses the state $|\varphi\rangle$ in this new basis. Because of the constraint \eqref{HS} the squared singular values, that is the eigenvalues of $GG^\dag$, are called Hilbert-Schmidt eigenvalues. They allow to quantify the degree of entanglement of 
$|\varphi\rangle$ by looking at the smallest (or largest) eigenvalue only. It distribution is known to be universal, agreeing with that of the Wishart-Laguerre ensemble without constraint \cite{Chen}. 
In contrast, when the Hamiltonians are non-Hermitian, we keep the bases of left and right eigenvectors instead, as typically operators acting on $|\varphi\rangle$ will yield complex eigenvalues. We then look at a Schur decomposition of a rectangular matrix $G$, and the resulting correlation functions of complex eigenvalues of $G$ in such a single FTE given by \eqref{HS}. 
To date only the spectral density of an $N\times N$ square matrix $G$ was computed at finite-$N$ \cite{DLC}. 
Our first goal is to extend this to arbitrary $k$-point correlation functions for rectangular matrices, the microcanonical version of the so-called induced Ginibre ensemble, c.f. \cite{Fischmann}. The challenge will then be to compute the distribution of the smallest eigenvalue in radial ordering.
We will also consider a single FTE with $G$ being normal, which is a priori  different.

The open question of local universality and of global higher order $k$-point functions was formulated in the second edition of \cite{Mehta:2004} for Hermitian FTE, that is their agreement with the classical, canonical Gaussian ensembles of RMT in the large-$N$ limit. It was shown in \cite{ACMV1,ACMV2} that more general FTE, where the trace of a polynomial of a Hermitian matrix is fixed, lead to the same limiting macroscopic spectral density as the corresponding canonical ensemble \cite{ACMV1}.
The finite-$N$ results of \cite{ACMV1} were confirmed independently in \cite{DLC} for the standard FTE. In contrast, the macroscopic, connected two-point \cite{ACMV2} and connected $k$-point functions \cite{AV} were shown to be non-universal. 
On the microscopic level, however,  heuristic arguments were given in \cite{AV} that local universality in the bulk persists for general FTE. This was made rigorous much later for standard FTE in \cite{GG}. For local universality in standard FTE at the soft edge see \cite{LiuZ}.
While all the quoted results apply to real eigenvalues in Hermitian FTE, very recently universality was proven for the complex eigenvalues of a  generalised FTE in the bulk of the spectrum, both at strong and weak non-Hermiticity \cite{ACV}.
While it is known that in the bulk the rate of convergence is exponentially fast for the Ginibre ensemble, see e.g. \cite{TV}, it is an open mathematical question if this property persists for other non-Gaussian ensemble. The finite-$N$ solution of the FTE investigated here will be an ideal testing ground for that question.

Products of random matrices have seen a very rapid development in the past five years. This is due to the observation that both, complex eigenvalues \cite{ABu} and singular values \cite{AKW,AIK} follow determinantal point processes at finite matrix and product size. Here, we will focus on the complex eigenvalues of complex matrices as they are simpler, and refer to the recent review \cite{AIp} for further literature, in particular on the singular values. 
We do not consider products of normal matrices here as their statistics is a difficult open problem, even without constraint.
The complex eigenvalues of products also enjoy an interpretation as a two-dimensional Coulomb gas, see \cite{PeterCoulomb} for a recent work regarding the respective interpretation of the induced Ginibre ensemble and products of Ginibre ensembles.
Furthermore, complex eigenvalues also play a role in the stability analysis of chaotic dynamical systems as initiated by \cite{FK}. They  lead to the so-called stability exponents, where the infinite product limit $M\to\infty$ is studied. In \cite{ABK} the agreement between the stability exponents, and the Lyapunov exponents based on singular values was found, including their variances, c.f. \cite{Peter1,Reddy}. In \cite{KK2}  the relation between complex eigenvalues and singular values valid already for finite products  of finitely many random matrices was found. 
Using classical harmonic analysis  a bijection was constructed in \cite{KK}, including more general classes of determinantal point processes of complex eigenvalues and singular values that go beyond products of random matrices. To date 
the complex eigenvalues statistics has been determined for finite matrix size for products of Gaussian random matrices from the complex, quaternion and real Ginibre ensembles \cite{ABu,Ipsen,FI}, inverse complex Ginibre \cite{Adhikari} and truncated unitary matrices \cite{Adhikari,IK,ABKN}, as well as for more general P\'olya ensembles defined in \cite{KK,FKK}. 
An immediate question regarding complex eigenvalue statistics of products of random matrices is that of the universality of the newly found classes on the global scale \cite{Burda,Burda2,GT}, on local scale at the origin \cite{ABu,ABKN}, as well as for the Gaussianity of the spectrum of the stability exponents \cite{ABK} relevant for chaotic dynamical systems. 
For a single ensemble of complex non-Hermitian matrices without constraint, universality has been studied rigorously by several authors \cite{ameur,berman,TV,Hedenmalm}. Adding a constraint, the ensembles become non-determinantal, and universality still holds \cite{ACV}.
What can be said about the universality of products of non-Gaussian random matrices? Very recently this problem has been addressed for Wigner matrices \cite{KRV}, however, the origin of the spectrum had to be spared out for technical reasons. 
Our study of multiplying FTE and induced Ginibre ensembles provides  an example for such a product, that can be explicitly solved at finite product and matrix size.
This serves  as a starting point to deepen our understanding of universality.

The remainder of the paper is organised as follows. In Section \ref{FTGin} we solve a single FTE, comparing the constraint on induced Ginibre and normal matrices. Although the partition function, joint density and resulting Coulomb gas of their complex eigenvalues are shown to be different in Subsection \ref{M1Zj}, we show in Subsection \ref{densitysect} that all their $k$-point correlation functions agree at finite-$N$, up to a trivial rescaling. They are obtained from an inverse Laplace transform of the well known induced Ginibre and normal Gaussian ensemble, respectively, which are known to agree.  For the distribution of the radii we show that these are no longer independent random variables, due to the constraint.
In  Section \ref{prodFT} we consider the product of $m$ induced Ginibre ensembles with constraint and $M-m$ without, with its joint density given in Subsection \ref{jpdfY}. 
In order to perform an $m$-fold inverse Laplace transform that is necessary here, we first solve for all $k$-point correlation functions of the product of $M$ independent induced Ginibre matrices with different variances in Subsection \ref{specY}. Due to the simple dependence on the variances and properties of Meijer G-functions, the spectral density can be explicitly computed for arbitrary fixed $m$ and $M$.  
We give heuristic arguments for the universality of this spectral density in the large-$N$ limit, and for the stability exponents at large $M$.
In Section \ref{conclusio} we summarise our findings. 
Several more technical identities are presented the two appendices \ref{AppA} and \ref{AppMeijer}.

\sect{Single fixed trace induced Ginibre and normal ensemble}
\label{FTGin}

In this section 
we introduce and solve two ensembles of random matrices $G$ at finite matrix size, with the constraint of fixing $\Tr GG^\dag$ by a delta-function.
The first ensemble is the induced Ginibre ensemble named in \cite{Fischmann} this way. It is equivalent to consider a rectangular matrix $\tilde{G}$ 
of size $N\times(N+\nu)$ for $\nu\in \mathbb{N}$, see also \cite{A01} for an earlier work. 
The second ensemble consists of normal random matrices $\G$ with $[\G,\G^\dag]=0$. 
Without the fixed trace constraint the distinction between Ginibre and Gaussian normal matrices is immaterial on the level of the joint density of complex eigenvalues, as we will recall below.  
This is due to the decoupling of the upper triangular degrees of freedom in the Schur decomposition of matrix $G$.

In contrast,  the introduction of a constraint will make these two ensembles distinct on the level of joint densities,  as it is shown in Subsection \ref{M1Zj}. 
It is easy to see that the ensembles with and without fixed trace constraint are simply related through a Laplace transform. 
Based on the known solution of the two ensembles without constraint, the inversion of this transformation will be our main tool of investigation.
Our results of Subsection \ref{densitysect} extend previous findings of \cite{DLC} for the spectral density of the fixed trace Ginibre ensemble at $\nu=0$ to arbitrary $k$-point density correlation functions at $\nu>-1$, and to the corresponding results for the fixed trace normal ensemble not considered in \cite{DLC}. 
We will also compute the gap probability at the origin.

\subsection{Partition functions, joint densities and Coulomb gases}\label{M1Zj}

The partition function of the fixed trace induced Ginibre ensemble is defined as 
\be
\Zd(s)\equiv \int[dG]\det[GG^\dag]^\nu\ \delta\left(s-\Tr GG^\dag\right)\ ,\ \ s>0\ , 
\label{Zds}
\ee
where $G\neq G^\dag$ is a non-Hermitian matrix of size $N\times N$ with complex elements, $\nu>-1$ is a real parameter, and the integration is over the flat measure of the real and imaginary parts of all independent matrix elements of $G$.
For integer $\nu$ an alternative representation exists as an integral over a rectangular matrix $\tilde{G}$ of size $N\times(N+\nu)$, without the determinant in the integrand, which can be seen from \cite{A01,Fischmann}.
It is clear that by rescaling the matrix $G\to\sqrt{s}G$ we could factor out the dependence on the parameter $s$. 
However, we will keep this parameter inside the delta-function, in order to 
relate this ensemble  to the induced Ginibre ensemble, defined in \eqref{Ztrafo} below, using the Laplace transform. For this reason we will keep track of all $s$-dependent normalisation constants.

The joint density of complex eigenvalues $z_{j=1,\ldots,N}$ of matrix $G$ from the ensemble \eqref{Zds} can be obtained after making the standard Schur decomposition
\be
G=U(Z+T)U^\dag\ .
\label{Schur}
\ee
Here, $Z=\mbox{diag}(z_1,\ldots,z_N)$ contains the complex eigenvalues, $T$ is a strictly upper triangular matrix with complex elements, thus having $N(N-1)$ real independent degrees of freedom, and $U$ is a unitary matrix from the following coset $U\in U(N)/U(1)^N$. 
For simplicity we will choose $N>1$ in all the following, as for $N=1$ the matrix $T$ is absent and the single eigenvalue is then fixed through the constraint. 
The Jacobian for the transformation (\ref{Schur}) computed in \cite{Ginibre}
contains the Vandermonde determinant 
\be
\Delta_N(Z)\equiv \prod_{1\leq i<j\leq N}(z_j-z_i) = \det_{1\leq i<j\leq N}\left[z_i^{j-1}\right]\ ,
\label{Vandermonde}
\ee
and we thus obtain for eq. (\ref{Zds}) 
\bea
\Zd(s)&=& \int[dU]\int[dT]\prod_{j=1}^N\int_{\mathbb C}d^2z_j|z_j|^{2\nu}
\ |\Delta_N(Z)|^2\ \delta\left(s-\Tr ZZ^*-\Tr TT^\dag \right)
\nn\\
&=& V_N \int d\Omega_{N(N-1)} \int_0^\infty dRR^{N(N-1)-1}
\prod_{j=1}^N\int_{\mathbb C}d^2z_j|z_j|^{2\nu}\ 
|\Delta_N(Z)|^2\ \delta\left(s-\sum_{l=1}^N|z_l|^2-R^2\right)\nn\\ 
&=& V_N\frac{\pi^{\frac{N(N-1)}{2}}}{\Gamma\left(\frac{N(N-1)}{2}\right)}
\prod_{j=1}^N\int_{\mathbb C}d^2z_j |z_j|^{2\nu}
\left(s- \sum_{l=1}^N|z_l|^2\right)^{\frac{N(N-1)}{2}-1}
|\Delta_N(Z)|^2\ \Theta\left(s-\sum_{l=1}^N|z_l|^2\right).\nn\\
&\equiv& \prod_{j=1}^N\int_{\mathbb C}d^2z_j P_{\delta,\nu}(z_1,\ldots,z_N;s)\ .
\label{Zdsev}
\eea
In the last line we 
have defined 
the (unnormalised) joint density $P_{\delta,\nu}(z_1,\ldots,z_N;s)$
of complex eigenvalues 
of our fixed trace ensemble \eqref{Zds}. It can be interpreted as a two-dimensional Coulomb gas at the particular inverse temperature $\beta=2$, by writing the integrand of the joint density from \eqref{Zdsev} as follows: 
\begin{equation}
\exp\left[2\nu\sum_{i=1}^N\log|z_i| +\sum_{j\neq i}\log|z_j-z_i|+\frac{N(N-1)-2}{2}\log\left[s-\sum_{l=1}^N|z_l|^2\right]
\right] \Theta\left(s-\sum_{l=1}^N|z_l|^2\right).
\label{CoulombFT}
\end{equation}
In addition to the standard two-dimensional Coulomb interaction in the first two terms in the exponent, with the first term coming from $2\nu$ charges at the origin, we have a collective potential in the last term of the exponent that depends on all particles and does not factorise. Together with the Heaviside constraint it leads to the confinement of all charges into a ball of squared radius $s$.

Furthermore, in \eqref{Zdsev} we have used
\be
V_N\equiv \int [dU]= \frac{\pi^{\frac{N(N-1)}{2}}}{\prod_{j=0}^Nj!}\ ,
\label{vol}
\ee
the volume of the coset integral, and the surface of the sphere of dimension $n$
\be
\int d\Omega_{n}=\frac{2\pi^{n/2}}{\Gamma(n/2)}\label{sphere}\ .
\ee
The latter is encountered when, following the ideas of \cite{ACMV1}, we change to polar coordinates $(R,\Omega_{N(N-1)})$ for the integral over matrix $T$. We then interpret $\Tr TT^\dag$ as the squared length of a vector of $N(N-1)$ real dimensions. 
This integral transforms the delta-function constraint into a step function constraint, containing the Heaviside function $\Theta(x)$. 
Due to this constraint the integrals are cut off and converge.

In comparison the ensemble of induced Gaussian normal matrices $\G$ with a fixed trace constraint that we label by superscript ${\cal N}$
is much simpler. It is formally defined as 
\bea
\Zdn(s)&\equiv& \int[d\G]\det[\G\G^\dag]^\nu\ \delta\left(s-\Tr \G\G^\dag\right)\ ,\ \ s>0\ , \label{Zdns}\\
 &=&\int[dU]\prod_{j=1}^N\int_{\mathbb C}d^2z_j|z_j|^{2\nu}
\ |\Delta_N(Z)|^2\ \delta\left(s-\sum_{l=1}^N|z_l|^2\right)\nn\\
&\equiv&\prod_{j=1}^N\int_{\mathbb C}d^2z_j P_{\delta,\nu}^{\mathcal{N}}(z_1,\ldots,z_N;s)\ ,
\label{Zdnsev}
\eea
where we could in fact use the second line as the definition, together with 
\be
\G=UZU^\dag\ .
\label{normal}
\ee
Also here the $s$-dependence could be scaled into a prefactor, by rescaling $\G\to\sqrt{s}\G$. 
As in the previous ensemble $Z=\mbox{diag}(z_1,\ldots,z_N)$ denotes the complex eigenvalues of $\G$ and 
we have $U\in U(N)/U(1)^N$. Because of the normality constraint $[\G,\G^\dag]=0$ and thus the dependence amongst the matrix elements of $\G$,  the initial measure $[d\G]$ is characterised by the second line in \eqref{Zdns}. Rewriting the integrand of the joint density \eqref{Zdnsev} as 
\begin{equation}
\exp\left[2\nu\sum_{i=1}^N\log|z_i| +\sum_{j\neq i}\log|z_j-z_i| \right] \delta\left(s-\sum_{l=1}^N|z_l|^2\right).
\label{CoulombFTN}
\end{equation}
we see that this ensemble represents a Coulomb gas different from \eqref{CoulombFT}, without a confining potential and a hard constraint.
In comparison, on the level of eigenvalues the previous ensemble \eqref{CoulombFT} looks more like a restricted trace  ensemble, see \cite{ACMV1,ACMV2} for a comparison between two such FTE of Hermitian matrices.
Let us emphasise here that the two joint densities of complex eigenvalues of our FTE, eqs. (\ref{Zdsev}) and (\ref{Zdnsev}), are {\it different} functions of the complex eigenvalues $\{z_1,\ldots,z_N\}$. 
Both are non-determinantal 
and coupled in a non-trivial way beyond the standard Vandermonde determinant, the Coulomb repulsion in two dimensions. Furthermore, due to the delta-function  in  the normal ensemble \eqref{Zdnsev}
its joint density $P_{\delta,\nu}^{\mathcal{N}}(z_1,\ldots,z_N;s)$ actually only depends on $N-1$ complex eigenvalues. 

As the next result of this subsection we can use the following relation to the ordinary induced Ginibre and normal ensemble to compute the integrals in eqs. (\ref{Zdsev}) and (\ref{Zdnsev}) as functions of $s$. Applying the Laplace transform,
\be
{\cal L}\{f(s)\}(t)= \int_0^\infty ds\ \e^{-ts}f(s) = F(t)\ ,
\ee
to eqs. (\ref{Zds}) and (\ref{Zdns})
immediately leads to the following relations:
\bea
{\cal L}\{\Zd(s)\}(t)&=& \int[dG]\det[GG^\dag]^\nu\exp[-t\,\Tr GG^\dag]\equiv \Z(t)
\ ,\ \ t>0\ ,
\label{Ztrafo}\\
{\cal L}\{\Zdn(s)\}(t)&=& \int[d\G]\det[\G\G^\dag]^\nu\exp[-t\,\Tr \G\G^\dag]
\ \ \equiv \Zn(t)\ ,\ \ t>0\ .
\label{Zntrafo}
\eea
Here, we define $\Z(t)$ and $\Zn(t)$ as the induced Ginibre and the induced Gaussian normal ensemble, respectively, which are well defined for $t>0$. Their joint densities and 
normalisations as functions of the (inverse) variance parameter $t$ read as follows. For the induced Ginibre ensemble we have
\bea
 \Z(t)&=& \int[dU]\int[dT]\exp[-t\,\Tr TT^\dag]
\prod_{j=1}^N\int_{\mathbb C}d^2z_j |z_j|^{2\nu}
\exp\left[- t\sum_{l=1}^N|z_l|^2\right]
|\Delta_N(Z)|^2\nn\\
&=&V_N \left(\frac{\pi}{t}\right)^{\frac{N(N-1)}{2}}\prod_{j=1}^N\int_{\mathbb C}d^2z_j |z_j|^{2\nu}
\exp\left[- t\sum_{l=1}^N|z_l|^2\right]
|\Delta_N(Z)|^2\nn\\
&\equiv& \prod_{j=1}^N\int_{\mathbb C}d^2z_j P_{\nu}(z_1,\ldots,z_N;t)
\label{ZGev}\\
&=& V_N \left(\frac{\pi}{t}\right)^{\frac{N(N-1)}{2}} N!\prod_{j=0}^{N-1}\frac{\pi\Gamma(j+1+\nu)}{t^{j+1+\nu}}
=
V_N \frac{\pi^{\frac{N(N+1)}{2}}}{t^{N^2+N\nu}}
N!\prod_{j=0}^{N-1}\Gamma(j+1+\nu)\ .
\label{ZGt}
\eea
In the last line we have used that the $N$ complex eigenvalue integrals 
can be computed in terms of the corresponding norms \eqref{norms} of orthogonal polynomials in the complex plane, to be introduced in the next Subsection \ref{densitysect}. 
The $t$-dependence agrees with what would be obtained from rescaling the matrix $\sqrt{t}G\to G$ in the definition \eqref{Ztrafo}.

For the induced Gaussian normal ensemble we obtain similarly 
\bea
 \Zn(t)&=& \int[dU]
\prod_{j=1}^N\int_{\mathbb C}d^2z_j |z_j|^{2\nu}
\exp\left[- t\sum_{l=1}^N|z_l|^2\right]
|\Delta_N(Z)|^2\nn\\
&\equiv& \prod_{j=1}^N\int_{\mathbb C}d^2z_j P_{\nu}^{\mathcal{N}}(z_1,\ldots,z_N;t)\ \ \ \ \ \ 
\label{Znev}\\
&=&  V_NN!\prod_{j=0}^{N-1}\frac{\pi\Gamma(j+1+\nu)}{t^{j+1+\nu}}
=
V_N \frac{\pi^N}{t^{\frac{N(N+1)}{2}+N\nu}}
N!\prod_{j=0}^{N-1}\Gamma(j+1+\nu)\ .
\label{Znt}
\eea
Here, the two joint densities eqs. (\ref{ZGev}) and (\ref{Znev}) {\it agree} as functions of the complex eigenvalues $z_j$, up to a $t$-dependent constant.  They represent 
the same Coulomb gas, obtained when writing the integrands of eqs. \eqref{ZGev} and \eqref{Znev} as 
\begin{equation}
\exp\left[2\nu\sum_{i=1}^N\log|z_i| +\sum_{j\neq i}\log|z_j-z_i|-t\sum_{l=1}|z_l|^2
\right].
\label{Coulomb}
\end{equation}
Compared to eqs. \eqref{CoulombFT} and \eqref{CoulombFTN} there is an additional confining Gaussian potential for each particle, apart from the constraint that is missing here.
The corresponding determinantal point processes \eqref{Coulomb} can be solved using the orthogonal polynomial technique in the complex plane, as we will recall in the next Subsection \ref{densitysect}.
Also here the dependence on the (inverse) variance parameter $t$ could be scaled out into a pre-factor, by redefining the random matrices $G$ and $\G$. For the map to the corresponding FTE it is more transparent to keep $t$ in the exponent though.
Applying the elementary inverse Laplace transform \cite{Grad},
\be
{\cal L}^{-1}\left\{\frac{1}{t^n}\right\}(s)=\frac{s^{n-1}}{\Gamma(n)}\Theta(s)\ ,\ \ n>0\ ,
\label{Laplace1}
\ee
to eqs. (\ref{Ztrafo}) and (\ref{Zntrafo}), together with the normalisations (\ref{ZGt}) and (\ref{Znt}) we obtain immediately 
\bea
\Zd(s)&=&{\cal L}^{-1}\{\Z(t)\}(s)=V_N \pi^{\frac{N(N+1)}{2}}\ \frac{N!\prod_{j=0}^{N-1}\Gamma(j+1+\nu)}{\Gamma(N^2+N\nu)}\ s^{N^2+N\nu-1}
\ , \ \ s>0\ ,
\label{Zdt}\\
\Zdn(s)&=&{\cal L}^{-1}\{\Zn(t)\}(s)=V_N\pi^N\  \frac{N!\prod_{j=0}^{N-1}\Gamma(j+1+\nu)}{\Gamma(\frac{N(N+1)}{2}+N\nu)}\ s^{\frac{N(N+1)}{2}+N\nu-1}\ , \ \ s>0\ ,
\label{Zdnt}
\eea
Here and in the following we omit the Heaviside function $\Theta(s)$, as by definition $s>0$. Also here the $s$-dependence is as expected from rescaling of the matrices in the definitions \eqref{Zds} and \eqref{Zdnsev}.
The leading order $N$-dependence of the two ensembles is different and can be related by replacing $N^2$ by $N(N-1)/2$, up factors of $\pi$. We will find the same relation between the correlation functions of the two ensembles in the next subsection, despite the different forms of their joint densities. 

We note that when integrating out all angles in the joint density \eqref{ZGev} and considering the resulting distribution 
$\mathcal{P}_{\nu}(r_1,\ldots,r_N;t)$
of the radii $r_j$ only, with  $z_j=r_j\e^{i\phi_j}$ for $j=1,\ldots,N$, 
the induced Ginibre and Gaussian normal ensemble simplify considerablly: 
\bea
\prod_{j=1}^N\int_0^\infty dr_j r_j\int_0^{2\pi}d\phi_j P_{\nu}(z_1,\ldots,z_N;t)&\equiv&\prod_{j=1}^N\int_0^\infty dr_j \mathcal{P}_{\nu}(r_1,\ldots,r_N;t)
\label{PGradii}\\
&=& V_N \left(\frac{\pi}{t}\right)^{\frac{N(N-1)}{2}}(2\pi)^N\prod_{j=1}^N\int_0^\infty dr_j \e^{-t r_j^2}\mbox{Per}[r_k^{2(l+\nu-1)}]_{k,l=1}^N\ .\nn\\
\label{PerG}
\eea
The same holds for the induced Gaussian normal ensemble, without the $t$-dependent prefactor.
Here, we have introduced the permanent of a matrix $A_{k,l}$ as per$[A]=\sum_{\sigma\in S_N}\prod_{l=1}^N A_{l,\sigma(l)}$. It differs from the determinant by the absence of the sign of the permutations $\sigma$. Eq. \eqref{PerG} can be easily seen by Laplace expanding the two Vandermonde determinants \eqref{Vandermonde}, and using that the monic polynomials $z^k$ are orthogonal with respect to angular integration, $\int_0^{2\pi}d\phi z^k z^{*\,l}\sim \delta_{k,l}$.
The statement \eqref{PerG} is equivalent to the fact that the $r_j$ become independent random variables \cite{Kostlan,AIS}. Because the joint densities of the induced Ginibre and Gaussian normal ensemble agree, \eqref{PerG} holds for both ensembles. 

Imposing the fixed trace constraint  spoils this interpretation as independent random variables, due to the remaining constraint. Defining the joint density of radii $\mathcal{P}_{\delta,\nu}(r_1,\ldots,r_N;t)$ as in \eqref{PGradii} by integrating out the angles in \eqref{Zdsev}, we have for fixed trace induced Ginibre ensemble
\be
\mathcal{P}_{\delta,\nu}(r_1,\ldots,r_N;s)=
V_N\frac{\pi^{\frac{N(N-1)}{2}}}{\Gamma\left(\frac{N(N-1)}{2}\right)}
(2\pi)^N
\mbox{Per}\left[r_k^{2(l+\nu-1)}\right]_{k,l=1}^N
\left(s- \sum_{l=1}^Nr_l^2\right)^{\frac{N(N-1)}{2}-1}
\Theta\left(s-\sum_{l=1}^Nr_l^2\right).
\label{Perd}
\ee
Likewise, we obtain for the joint density of the radii of the ensemble of induced Gaussian normal matrices with fixed trace constraint
\be
\mathcal{P}^{\mathcal{N}}_{\delta,\nu}(r_1,\ldots,r_N;s)=
V_N
(2\pi)^N
\mbox{Per}[r_k^{2(l+\nu-1)}]_{k,l=1}^N
\delta\left(s-\sum_{l=1}^Nr_l^2\right).
\label{PerdN}
\ee
In both eqs. \eqref{Perd} and \eqref{PerdN} the independence of the radii is lost as they become coupled in the delta- or theta-constraint.

\subsection{Density $k$-point correlation functions and gap probability}\label{densitysect}

We turn to the $k$-point density correlation functions.
For FTE they are defined for $k=1,\ldots,N$ as
\bea
R_{\delta,\nu}^{(k)}(z_1,\ldots,z_k;s) &\equiv& \frac{N!}{(N-k)!}\frac{1}{\Zd(s)}
\int_{\mathbb C}d^2z_{k+1}\ldots \int_{\mathbb C}d^2z_N 
P_{\delta,\nu}(z_1,\ldots,z_N;s)
\ ,\label{Rkdef}\\
&=& \frac{1}{(N-k)!}\frac{\Gamma(N^2+N\nu)}{\pi^N\Gamma\left(\frac{N(N-1)}{2}\right)}
\frac{s^{-N^2-N\nu+1}}{\prod_{j=0}^{N-1}\Gamma(j+1+\nu)}
\int_{\mathbb C}d^2z_{k+1}\ldots \int_{\mathbb C}d^2z_N \nn\\
&&\times \prod_{j=1}^N|z_j|^{2\nu}
\left(s- \sum_{l=1}^N|z_l|^2\right)^{\frac{N(N-1)}{2}-1}
|\Delta_N(Z)|^2\ \Theta\left(s-\sum_{l=1}^N|z_l|^2\right),
\label{Rkd}
\eea
where we integrate the joint density from \eqref{Zdsev}, normalised with respect to the partition function \eqref{Zdt}, over the remaining eigenvalues from $z_{k+1}$ to $z_N$.
For $k=N$ there are no integrals and the $k$-point function equals the joint density times $N!$. 
The $k$-point densities for all other ensembles are defined according to \eqref{Rkdef}, too\footnote{Because the joint density in the normal FTE depends on $N-1$ variables we have to consider $k\leq N-1$ there.}. 

Because of the lack of determinantal structure there is no obvious way to directly compute the integrals in \eqref{Rkd}. For the density a similar strategy as for computing the partition function in eq. (\ref{Zdsev}) is available, by going to polar coordinates $(R,\Omega_{2N-2})$ for the integrated $N-1$ complex eigenvalues, and comparing the remaining angular integral to the known density of the induced Ginibre ensemble. This calculation can be found in appendix \ref{alternative}, cf. \cite{Milan}. For increasing $k\geq2$ this strategy becomes rapidly cumbersome. 

Therefore, we will use instead the map to the induced Ginibre ensemble via Laplace transform, which can be applied to all $k$-point correlation functions. Inserting the first line of \eqref{Zdsev} for the joint density into definition (\ref{Rkdef}), clearly the Laplace transform applies to unnormalised $k$-point densities,
\bea
{\cal L}\left\{\Zd(s)R_{\delta,\nu}^{(k)}(z_1,\ldots,z_k;s)\right\}(t)&=&
\frac{N!}{(N-k)!}
\int_{\mathbb C}d^2z_{k+1}\ldots \int_{\mathbb C}d^2z_N \nn\\
&&\times\int[dU] \int[dT] \prod_{j=1}^N|z_j|^{2\nu}
\ |\Delta_N(Z)|^2\ \exp[-t\,\Tr ZZ^*-t\,\Tr TT^\dag]
\nn\\
&=&
\frac{N!V_N\left(\frac{\pi}{t}\right)^{\frac{N(N+1)}{2}}}{(N-k)!}
\int_{\mathbb C}d^2z_{k+1}\ldots \int_{\mathbb C}d^2z_N 
\!\prod_{j=1}^N|z_j|^{2\nu}
\e^{-t|z_j|^2}
|\Delta_N(Z)|^2
\nn\\
&=& \Z(t)R_{\nu}^{(k)}(z_1,\ldots,z_k;t)\ .
\label{Rktrafo}
\eea
Here, we have compared with \eqref{ZGev}, to arrive at the unnormalised $k$-point correlation function of the induced Ginibre ensemble.
Following the theory of orthogonal polynomials in the complex plane \cite{Mehta:2004}, the latter is known explicitly, and performing the inverse Laplace transform of \eqref{Rktrafo} we can read off the $R_{\delta,\nu}^{(k)}(z_1,\ldots,z_k;s)$
for the fixed trace induced Ginibre ensemble, after dividing by $\Zd(s)$.

In order to prepare this step let us recall the results for the induced Ginibre ensemble, following \cite{Fischmann}.
Because in the definition of the $k$-point density correlation functions following from \eqref{Rkdef}, the normalisation constants, by which the joint densities \eqref{ZGev} and \eqref{Znev} of the induced Ginibre and induced Gaussian normal ensemble differ, drop out, and the two ensembles  agree for all $k$-point correlation functions. 
The weight function 
\be 
w(z;t)=|z|^{2\nu}\exp[-t|z|^2]
\label{Ginweight}
\ee
of the two ensembles is rotationally invariant and thus the monic orthogonal polynomials in the complex plane are trivially given by the monomials, $P_k(z)=z^k$. We find for the $t$-dependent (squared) norms $h_k(t)$  of these polynomials as
\be
\int_{\mathbb C}d^2z |z|^{2\nu}\e^{-t|z|^2} z^kz^{*\,l}=\delta_{k,l}h_k(t)
\ , \ \ 
h_k(t)=\pi\Gamma(k+1+\nu)t^{-k-1-\nu}\ .
\label{norms}
\ee
The kernel of orthonormalised polynomials thus reads
\be
K_{\nu}(z,u^*;t)=|zu|^{\nu}\exp\left[-\frac{t}{2}(|z|^2+|u|^2)\right]\sum_{l=0}^{N-1}
\frac{(zu^*)^l}{\pi\Gamma(l+1+\nu)}t^{l+1+\nu}\ ,
\label{kernel}
\ee
which is an explicit function of $t$.
The $k$-point density correlations functions of the induced Ginibre (and normal) ensemble are then given by the determinant of this kernel
\be
R_{\nu}^{(k)}(z_1,\ldots,z_k;t)=\det_{1\leq i,j\leq k}\left[ K_{\nu}(z_i,z_j^*;t)\right]=R_{\nu}^{(k)\cal N}(z_1,\ldots,z_k;t)\ .
\label{Rkdet}
\ee
In particular, for $k=1$ the spectral  density (1-point correlation function) is given by 
\be
R_{\nu}^{(1)}(z;t)=R_{\nu}^{(1)\cal N}(z;t)
=K_{\nu}(z,z^*;t)=|z|^{2\nu}\e^{-t|z|^2}\sum_{l=0}^{N-1}
\frac{|z|^{2l}}{\pi\Gamma(l+1+\nu)}t^{l+1+\nu}\ .
\label{Gdensity}
\ee
Its integral is normalised to $N$ in our convention \eqref{Rkdef}.

Finally, we compute the result for the partition functions of the induced Ginibre and normal ensemble \eqref{ZGt} and \eqref{Znt}, respectively. Following general theory of orthogonal polynomials \cite{Mehta:2004},
the integrals over the complex eigenvalues in (\ref{ZGev}) or \eqref{Znev}
are given by the product of the norms of the orthogonal polynomials times $N!$, 
\be
\Z(t)= V_N \left(\frac{\pi}{t}\right)^{\frac{N(N-1)}{2}} N!\prod_{j=0}^{N-1}h_j(t)
=\left(\frac{\pi}{t}\right)^{\frac{N(N-1)}{2}}\ \Zn(t)\ .
\label{Zprodh}
\ee
Together with \eqref{norms} this leads to eq.  (\ref{ZGt}), and analogously to eq. (\ref{Znt}).

In order to obtain the $k$-point correlation functions of our two FTE, we need to invert  
the Laplace transform \eqref{Rktrafo}, after inserting the right hand side from \eqref{Rkdet}. Here, we use 
another standard formula for the inverse Laplace transform:
\be
{\cal L}^{-1}\left\{\frac{1}{t^n}\ \e^{-t|z|^2}\right\}(s)=\frac{(s-|z|^2)^{n-1}}{\Gamma(n)}\Theta(s-|z|^2)\ ,\ \ n>0\ .
\label{Laplace2}
\ee
As an example let us first perform the inverse Laplace transform for the spectral density at $k=1$. From eqs. (\ref{Rktrafo}) and \eqref{Gdensity} we have 
\bea
\Zd(s)R_{\delta,\nu}^{(1)}(z;s)&=&
{\cal L}^{-1}\left\{\Z(t)R_{\nu}^{(1)}(z;t)\right\}(s)\nn\\
&=& {\cal L}^{-1}\left\{
V_N \frac{\pi^{\frac{N(N+1)}{2}}}{t^{N^2+N\nu}}
N!\prod_{j=0}^{N-1}\Gamma(j+1+\nu)
\sum_{l=0}^{N-1}
\frac{|z|^{2l+2\nu}\e^{-t|z|^2}}{\pi\Gamma(l+1+\nu)}t^{l+1+\nu}
\right\}(s)\nn\\
&=&
V_N \pi^{\frac{N(N+1)}{2}}N!\prod_{j=0}^{N-1}\Gamma(j+1+\nu)
\sum_{l=0}^{N-1}\frac{|z|^{2l+2\nu}(s-|z|^2)^{N^2+N\nu-l-\nu-2}\Theta(s-|z|^2)}{\pi\Gamma(l+1+\nu)
\Gamma(N^2+N\nu-l-\nu-1)}.\nn\\
\label{RdLaplace}
\eea
for the exponent we have $N^2+N\nu-l-1-\nu\geq N^2+N\nu-N-\nu=(N+\nu)(N-1)>0$. This is always satisfied for $N>1$.
After dividing by \eqref{Zdt},
the final answer for the spectral density of the fixed trace induced Ginibre ensemble thus reads: 
\bea
R_{\delta,\nu}^{(1)}(z;s)&=&\Gamma(N^2+N\nu) s^{-N^2-N\nu+1}
\sum_{l=0}^{N-1}\frac{|z|^{2l+2\nu}(s-|z|^2)^{N^2+N\nu-l-\nu-2}\Theta(s-|z|^2)}{\pi\Gamma(l+1+\nu)
\Gamma(N^2+N\nu-l-\nu-1)}\nn\\
&=&\frac{(N^2+N\nu-1)}{\pi s^{N^2+N\nu-1}}
\sum_{l=0}^{N-1}
{{N^2+N\nu-2}\choose{l+\nu}}
{|z|^{2l+2\nu}(s-|z|^2)^{N^2+N\nu-l-\nu-2}}\ 
\Theta(s-|z|^2).\ \ \ \ \ \ \ \ 
\label{Rdfinal}
\eea
In the second step we have written the terms under the sum as a binomial distribution. This sum could be expressed in terms of a linear combination of two hypergeometric functions $_2F_1$ which we do not display here.
Alternatively, the sum can be simplified to contain only powers in $|z|^2$ as shown in the Appendix \ref{alternative}. We arrive at
\bea
R_{\delta,\nu}^{(1)}(z;s)&=&\frac{(N^2+N\nu-1)}{\pi s}\left(1-\frac{|z|^2}{s}\right)^{N^2+N\nu-N-1} \Theta\left(1-\frac{|z|^2}{s}\right) \nn\\
&&\times
\sum_{m=0}^{N-1-\nu}\left( \frac{|z|^2}{s}\right)^{m+\nu}\sum_{l=0}^m {{N^2+N\nu-2}\choose{l+\nu}}{{-N+m+\nu}\choose{m-l}} .
\label{Rdfinal-alt}
\eea
Clearly $s$ times the density is a function of the combination $|z|^2/s$ only.
The second binomial coefficient is understood as 
\be
{{-N+m+\nu}\choose{m-l}}\!\!=\frac{(-N+m+\nu)(-N+m+\nu-1)\cdots\!(-N+l+1+\nu)}{(m-l)!}=(-1)^{m-l}{{N-l-\nu-1}\choose{m-l}}\!.
\ee
In the special case of
the fixed trace Ginibre ensemble with $\nu=0$ and $s=1$, the spectral density was previously computed 
by Delannay and LeCa\"er \cite{DLC}.
In this case for $\nu=0$ the second sum in \eqref{Rdfinal-alt} can be simplified, 
\be
\sum_{l=0}^m{{N^2-2}\choose{l}}{{-N+m}\choose{m-l}}
=\sum_{l=0}^m{{N^2-2}\choose{l}}{{N-l-1}\choose{m-l}}(-1)^{m-l}={{N^2-N-2+m}\choose{m}},
\ee
which follows from standard identities for binomial coefficients. This yields the form computed in \cite{DLC}.
\begin{figure}[h]
\centerline{\includegraphics[scale=0.6]{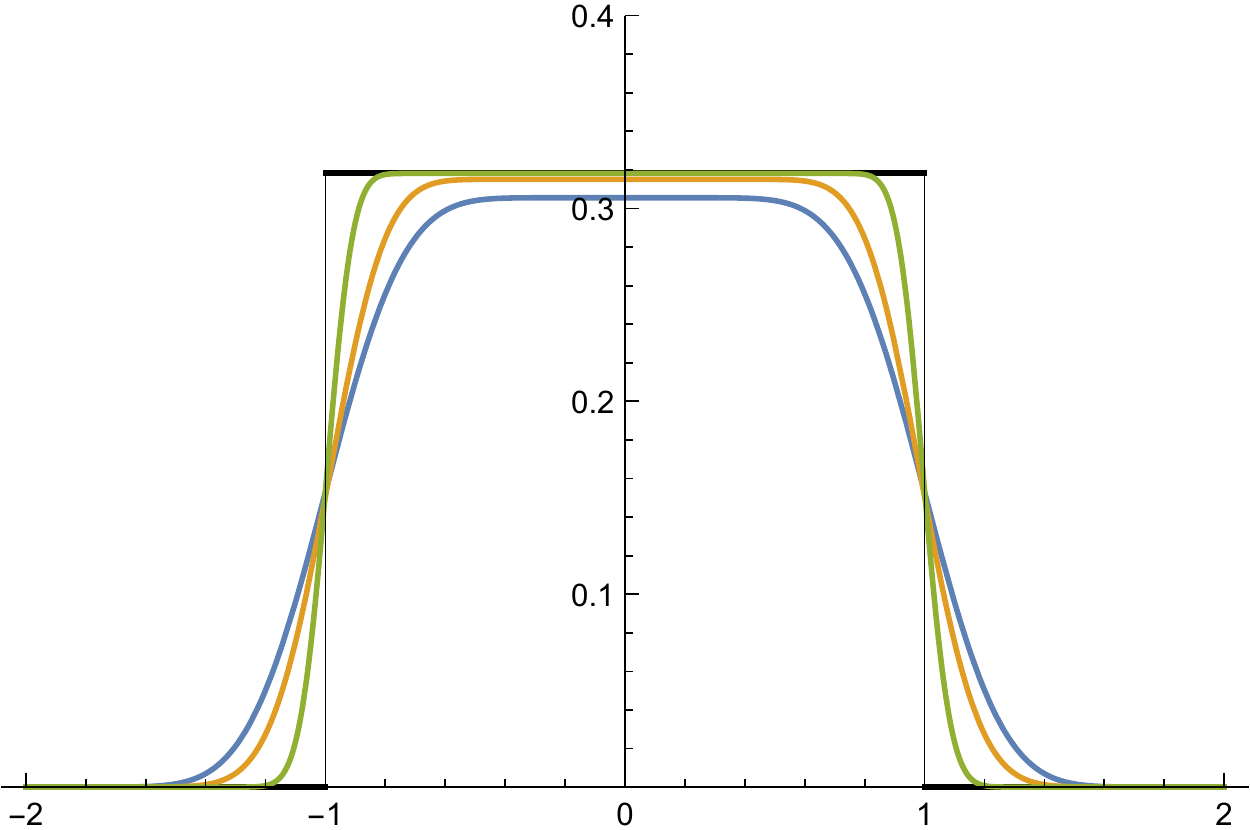}
\includegraphics[scale=0.6]{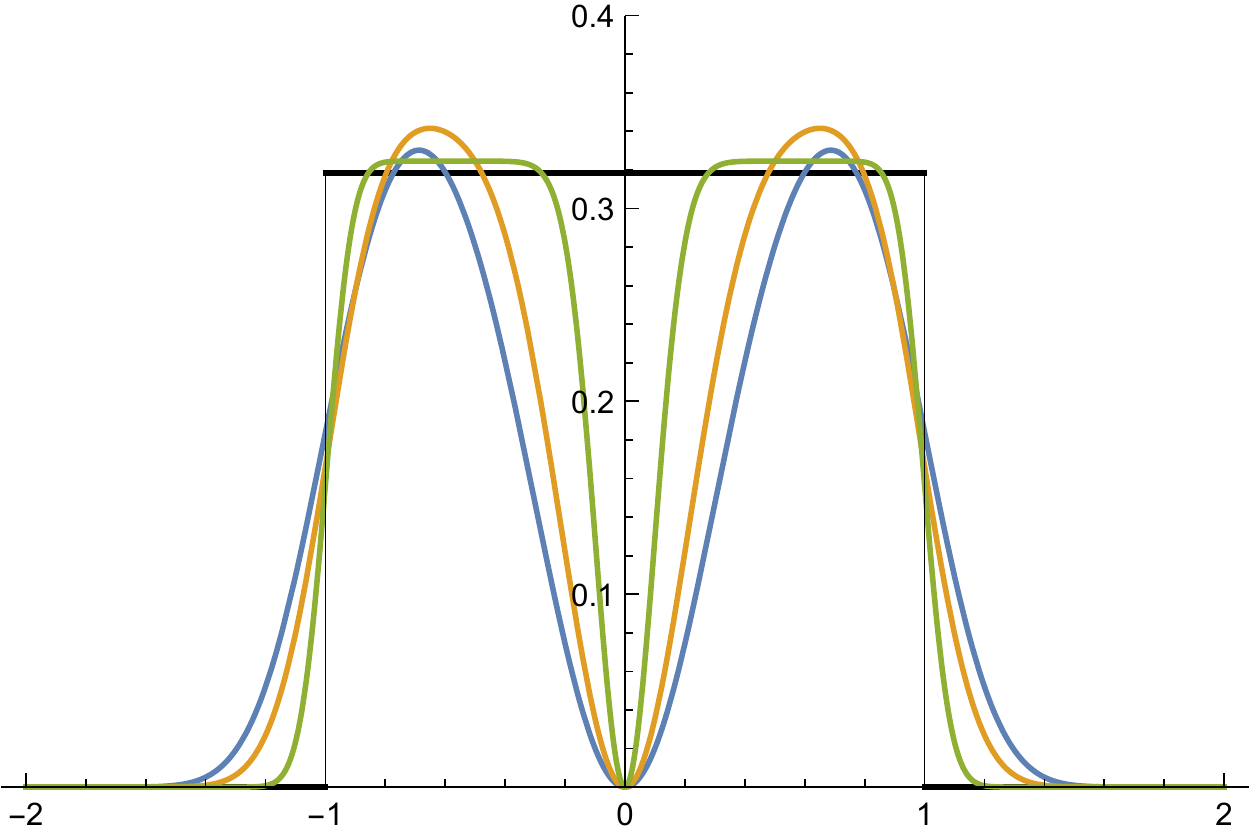}}
\centerline{\includegraphics[scale=0.6]{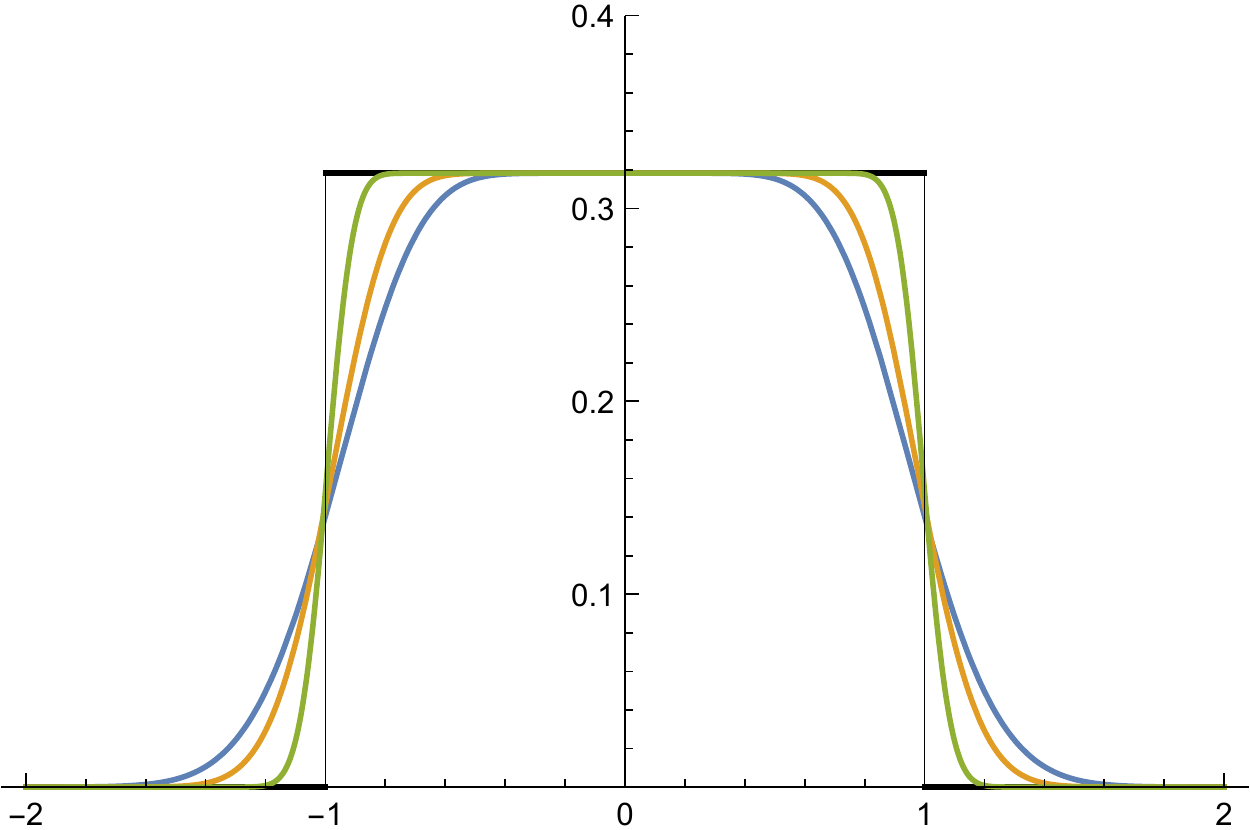}
\includegraphics[scale=0.6]{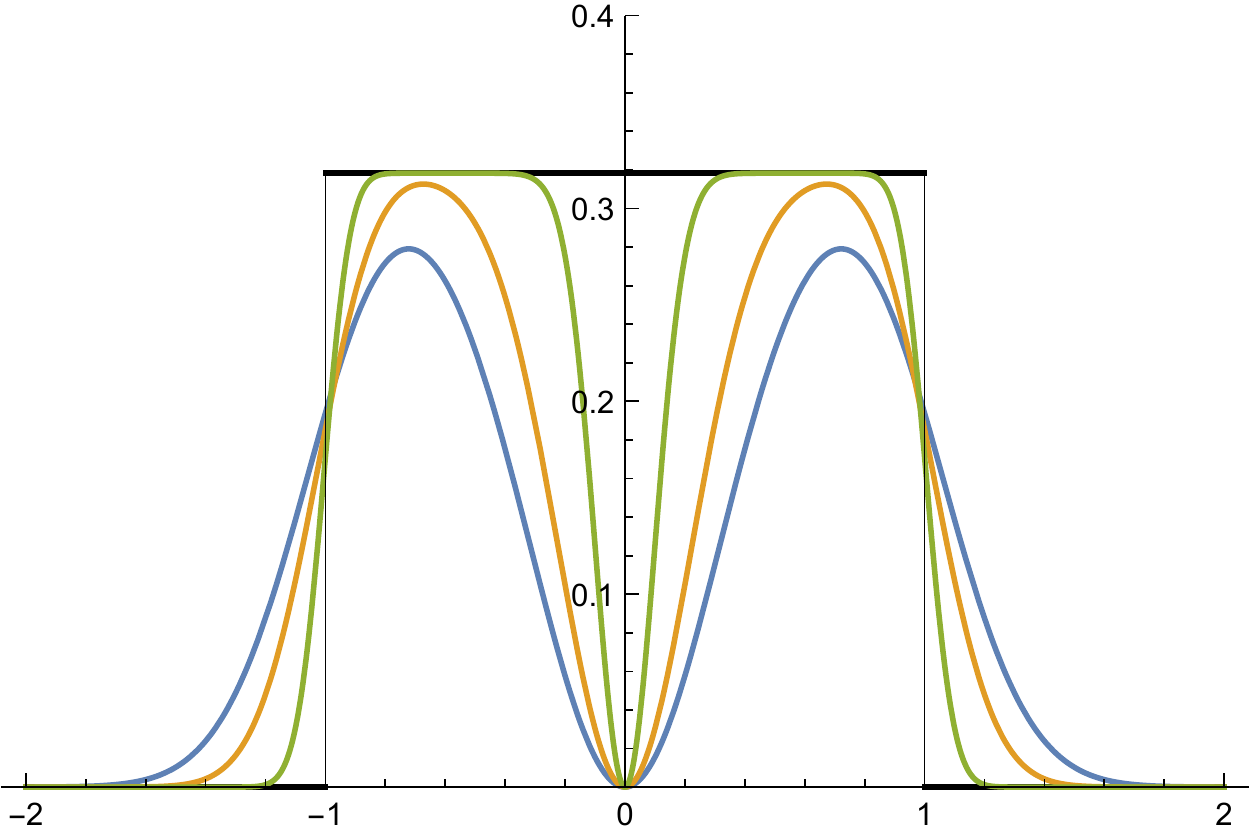}}
\caption{We plot a cut through the rescaled spectral density $R_{\delta,\nu}^{(1)}(z/\sqrt{N};s=1)/N^2$ 
of the fixed trace induced Ginibre ensemble from \eqref{Rdfinal}
for $z=x+i0$ against $x\in[-2,2]$, with $\nu=0$ (top left) and $\nu=1$ (top right). The different values of $N=5,10$, and $50$ correspond to the bottom blue, middle orange and top green line, respectively. 
 For comparison the scaled spectral density $R_{\nu}^{(1)}(z\sqrt{N};t=1)$ of the induced Ginibre ensemble from \eqref{Gdensity} is shown for $\nu=0$ (bottom left) and $\nu=1$ (bottom right), with the same parameter values for $N$.
 The limiting circular law is also shown in all plots (black line).} 
\label{densityM=1Gin-FT}
\end{figure}
The authors used different techniques, exploiting a map to so-called spherical ensembles and a combination of Laplace transform and polar coordinates. 
As mentioned already after \eqref{Rdfinal-alt},
the rescaling 
\be
s^{-1}R_{\delta,\nu}^{(1)}\left(z/\sqrt{s};s\right)= R_{\delta,\nu}^{(1)}(z;s=1)
\label{scale1}
\ee
makes the density independent of the constraint parameter $s$.

The spectral density for the fixed trace induced ensemble of normal matrices can be obtained in the same way, 
as derived in the Appendix \ref{FTn}. 
Despite the different structure of the joint densities of eigenvalues  (\ref{Zdsev}) and (\ref{Zdnsev}) and corresponding different Coulomb gases \eqref{CoulombFT} and \eqref{CoulombFTN},
the final answer for the spectral density only differs in replacing $N^2$ by $\frac{N(N+1)}{2}$ in \eqref{Rdfinal}, 
\bea
R_{\delta,\nu}^{(1)\,\cal N}(z;s)
&=&\frac{\frac{N(N+1)}{2}+N\nu-1}{\pi s^{\frac{N(N+1)}{2}+N\nu-1}}
\sum_{l=0}^{N-1}
{{\frac{N(N+1)}{2}+N\nu-2}\choose{l+\nu}}
{|z|^{2l+2\nu}(s-|z|^2)^{\frac{N(N+1)}{2}+N\nu-l-\nu-2}}
\nn\\
&&\times
\Theta(s-|z|^2)\ .
\label{Rdnfinal}
\eea
This clearly indicates that the fixed trace induced Ginibre and fixed trace induced Gaussian normal ensemble belong to the same universality class.
Consequently,  the density times $s$ is the same function of $|z|^2/s$, 
and thus the same rescaling as in \eqref{scale1} applies.

In Fig. \ref{densityM=1Gin-FT} we compare the spectral densities \eqref{Rdfinal} and \eqref{Gdensity}, with and without constraint at $\nu=0,1$.
Note however, that the rescaling has to be chosen differently for the fixed trace and induced Ginibre ensemble, so that they approach the same 
limiting curve. On the global scale it is given by the circular law, $\rho(z)=\frac{1}{\pi}\Theta(1-|z|)$, which is also plotted, cf. \cite{ACV}.
For the 
Ginibre ensemble it is well known that the rate of convergence in the bulk of the spectrum is exponentially fast \cite{TV}.
Therefore it is striking that in the bulk convergence is much slower for the fixed trace ensemble. This can be easily seen analytically for $\nu=0$ at the origin $z=0$ as an example. Here, the Ginibre density \eqref{Gdensity} already takes the limiting value $1/\pi$ at finite $N$, whereas for fixed trace the rescaled density \eqref{Rdfinal} approaches this value only algebraically as $(1-\frac{1}{N^2})/\pi$ 
(the rate obtained in \cite{ACV} was clearly not optimal). 
This shows that the rate of convergence for the Ginibre ensemble which is Gaussian is rather special.

In order to explain the different scaling in $N$ plotted in Fig. \ref{densityM=1Gin-FT},
we give a simple heuristic argument, relating the spectral densities \eqref{Rdfinal} and \eqref{Gdensity} with and without fixed trace constraint as follows: 
\bea
&&\lim_{N\gg1} \frac{1}{N^2} R_{\delta,\nu}^{(1)}\left(\frac{z}{\sqrt{N}};1\right)\nn\\
&&=\lim_{N\gg1}\! \frac{(N^2+N\nu-1)}{N^2}\!
\sum_{l=0}^{N-1}
\frac{(N^2+N\nu-2)!}{(N^2+N\nu-2-l-\nu)!}\frac{|z|^{2l+2\nu}}{N^{l+\nu}}
\frac{\left(1-\frac{N|z|^2}{N^2}\right)^{N^2+N\nu-l-\nu-2}}{\pi(l+\nu)!}
\Theta\left(1-\frac{|z|^2}{N}\right)
\nn\\
&&\simeq \sum_{l=0}^{N-1}
\frac{(N|z|^2)^{l+\nu} \e^{-N|z|^2}}{\pi(l+\nu)!} 
\ =\  R_{\nu}^{(1)}\left(z\sqrt{N};1\right)\ .
\label{scalingM=1}
\eea
To leading order the prefactor of the sum tends to unity. In the first factor under the sum the factorials cancel up to factors of $N$, that provide a scaling of $N|z|^2$ in the next factor.  
The third factor in the numerator is simply replaced by an exponential, leading approximately to the rescaled density of the Ginibre ensemble \eqref{Gdensity} (without Heaviside function).
The latter density is known to have compact support on the unit disc for large $N$ in this scaling. The density at finite $N\gg1$  gives rise to the different global and local scaling limits in the various large-$N$ regimes, cf. \cite{Mehta:2004}.\\

Next we turn to the general case and perform the inverse Laplace transform of eq. (\ref{Rktrafo}), in order to obtain all $k$-point density correlation functions. Our strategy is as follows: we will first Laplace expand the determinant of the kernel in eq. (\ref{Rkdet}) and then collect all $t$-dependence in terms of powers and exponentials to be transformed, 
\bea
\Zd(s)R_{\delta,\nu}^{(k)}(z_1,\ldots,z_k;s)&=&
{\cal L}^{-1}\left\{ \Z(t)\det_{1\leq i,j\leq k}\left[K_{\nu}(z_i,z_j^*;t)\right]\right\}(t)\nn\\
&=&{\cal L}^{-1}\left\{V_N \frac{\pi^{\frac{N(N+1)}{2}}}{t^{N^2+N\nu}}
N!\prod_{j=0}^{N-1}\Gamma(j+1+\nu)\sum_{\sigma\in S_k}(-1)^\sigma\prod_{j=1}^k
K_{\nu}(z_j,z_{\sigma(j)}^*;t)\right\}\!(s)\nn\\
\label{Rkinv}
\eea
with 
\be
\prod_{j=1}^k
K_{\nu}(z_j,z_{\sigma(j)}^*;t)=
\prod_{j=1}^k\left(
\sum_{l_j=0}^{N-1}
\frac{(z_jz_{\sigma(j)}^*)^{l_j}}{\pi\Gamma(l_j+1+\nu)}t^{l_j+\nu+1}
|z_j|^{2\nu}\e^{-t|z_j|^2}
\right).
\ee
The inverse Laplace transform of \eqref{Rkinv} can now be taken, using  (\ref{Laplace2}), under the condition that the total power in $1/t^n$, given by 
$n=N^2+N\nu-k(\nu+1)-\sum_{j=1}^kl_j$,
is positive. We obtain
\bea
&&\Zd(s)R_{\delta,\nu}^{(k)}(z_1,\ldots,z_k;s)\ =\nn\\
&=&V_N 
{\pi^{\frac{N(N+1)}{2}}}
N!\prod_{j=0}^{N-1}\Gamma(j+1+\nu)\sum_{\sigma\in S_k}(-1)^\sigma\prod_{j=1}^k
\left(\sum_{l_j=0}^{N-1}\frac{|z_j|^{2\nu}(z_jz_{\sigma(j)}^*)^{l_j}}{\pi\Gamma(l_j+1+\nu)}
\left(s-\sum_{i=1}^k|z_i|^2\right)^{-l_j-1-\nu}\right)\nn\\
&&\times \frac{(s-\sum_{i=1}^k|z_i|^2)^{N^2+N\nu-1}}{\Gamma\left(N^2+N\nu-\sum_{n=1}^kl_n-k(\nu+1)\right)}\ \Theta\left(s-\sum_{i=1}^k|z_i|^2\right).
\label{RkLaplace}
\eea
Because of  $n\geq N^2+N\nu-k(N+\nu)=N(N-k)+\nu(N-k)>0$, this condition holds for $N>k$ which we can always satisfy for any given fixed $k$ for large enough $N$. 
After dividing \eqref{RkLaplace} by \eqref{Zdt}, the final answer for the $k$-point density correlation functions reads
\bea
R_{\delta,\nu}^{(k)}(z_1,\ldots,z_k;s)&=&\frac{\Gamma(N^2+N\nu)\prod_{j=1}^k|z_j|^{2\nu}}{s^{N^2+N\nu-1}}
\left(s-\sum_{i=1}^k|z_i|^2\right)^{N^2+N\nu-1}\Theta\left(s-\sum_{i=1}^k|z_i|^2\right)
\nn\\
&&\times\sum_{\sigma\in S_k}(-1)^\sigma\prod_{j=1}^k
\left(\sum_{l_j=0}^{N-1}\frac{(z_jz_{\sigma(j)}^*)^{l_j}}{\pi\Gamma(l_j+1+\nu)\left(s-\sum_{i=1}^k|z_i|^2\right)^{l_j+1+\nu}}
\right)\nn\\
&&\times \frac{1}{\Gamma\left(N^2+N\nu-\sum_{n=1}^kl_n-k(\nu+1)\right)} .
\label{Rkdfinal}
\eea
It is the inverse Gamma-function in the last line that couples all sums in the product and that destroys the determinantal structure, that was present for the induced Ginibre (and normal) ensemble, see \eqref{Rkdet}.
It is not difficult to see that $s^k$ times the $k$-point function is a function of $|z|^2/s$ only, and that the following rescaling makes it independent of the constraint parameter:
\be
s^{-k}R_{\delta,\nu}^{(k)}\left(z_1/\sqrt{s},\ldots,z_k/\sqrt{s};s\right)=R_{\delta,\nu}^{(k)}(z_1,\ldots,z_k;s=1)\ .
\label{scalek}
\ee

The $k$-point correlation functions for the corresponding normal ensemble can be obtained in a similar way, see \eqref{RkdNfinal} in Appendix \ref{FTn}, to where we refer for the explicit answer. 
They only differ from \eqref{Rkdfinal} again by the replacement $N^2\to\frac{N(N+1)}{2}$. Consequently, 
the fixed trace induced Ginibre and fixed trace induced Gaussian normal ensemble belong to the same universality class. For Hermitian restricted and FTE such an agreement is only true in the large-$N$ limit, see \cite{ACMV1,ACMV2}.

For the $k$-point correlation functions of the FTE
it is more involved to see that in the large-$N$ limit the determinantal structure \eqref{Rkdet} is reestablished, and we refer to \cite{ACV} for a proof of this statement using different methods.

Last but not least we turn to the question of gap probability and the distribution of the smallest eigenvalue in radial ordering in FTE.
In the induced Ginibre and normal ensemble these can be computed easily, 
using the same orthogonality of the monic polynomials with respect to angular integration as in Subsection \ref{M1Zj} when discussing the distribution of the radii. Following \cite{Mehta:2004,APS} we define the probability to have no eigenvalues inside the disc of radius $x$ as
\begin{eqnarray}
E_\nu(x;t)&\equiv& \frac{1}{ \Z(t)}\int[dU]\int[dT]\exp[-t\,\Tr TT^\dag]
\prod_{j=1}^N\int_0^{2\pi}d\phi_i\int_x^\infty dr_j r_j^{2\nu} \exp\left[- tr_i^2\right]\ 
|\Delta_N(Z)|^2\ \ \ \ \ \ \ \ \label{Gapt}\\
&=& \frac{1}{ \Z(t)}V_N \left(\frac{\pi}{t}\right)^{\frac{N(N-1)}{2}}N!\det\left[
\int_0^{2\pi}d\phi e^{i(k-l)\phi}\int_x^\infty dr r^{1+2\nu+k+l}e^{-tr^2}
\right]_{k,l=0}^{N-1}
\nn\\
&=& \prod_{j=0}^{N-1}\frac{\Gamma(j+1;tx^2)}{\Gamma(j+1)}\ .
\label{Gaptresult}
\end{eqnarray}
Here, we have used the Andr\'eief integration formula, orthogonality and collected all factors to arrive at the product of Fredholm eigenvalues given in terms of the incomplete gamma function. The distribution of the smallest eigenvalue in radial ordering follows from differentiation. 
It is clear that the corresponding gap probability $E_{\delta,\nu}(x;t)$ for our FTE, defined as in \eqref{Gapt}, follows from a Laplace transform of 
the unnormalised quantity \eqref{Gapt},
as in \eqref{Rktrafo}: 
\begin{equation}
{\cal L}\left\{\Zd(s)E_{\delta,\nu}(x;s)\right\}(t)= \Z(t)E_{\delta,\nu}(x;t)\ ,
\end{equation}
and likewise for the normal ensemble.
Therefore,  the gap probability in our  FTE can be written as
\begin{equation}
E_{\delta,\nu}(x;s)=\Gamma(N^2+N\nu)s^{-N^2-N\nu+1}
\int_{\gamma-i\infty}^{\gamma+i\infty}\frac{dt}{2\pi i}e^{st} t^{-N^2-N\nu}\prod_{j=0}^{N-1}\frac{\Gamma(j+1;tx^2)}{\Gamma(j+1)},
\end{equation}
chosing the contour appropriately.
Here, we inserted  \eqref{Gaptresult}, and for the prefactor we used \eqref{Zdns} and \eqref{Znev}, canceling all common factors. The result is reminiscent but not identical to a Meijer G-function in the form \eqref{Meijer-def}, and it remains to be seen if it can be evaluated more explicitly.

\sect{Product of $m$ fixed trace and $M-m$ induced Ginibre matrices}
\label{prodFT}

In the previous Section \ref{FTGin} we  have solved the induced Ginibre ensemble with a fixed trace constraint for finite $N$, as well as its normal version. 
We are now prepared to study the complex eigenvalues of the product $Y_M$ of $M\geq m\geq1$ induced Ginibre matrices $G_{1},\ldots,G_{m}$ with a fixed trace constraint, times $M-m$ matrices $G_{m+1},\ldots,G_M$ from the induced Ginibre ensemble without constraint,
\be
Y_M\equiv G_1 \cdots G_m G_{m+1}\cdots G_M \ .
\label{Pidef}
\ee
Because the complex eigenvalue statistics of products of normal matrices without constraint is already a difficult open problem, and our main tool is the inverse Laplace transform of ensembles which are known, we do not consider normal matrices in this section at all.  
Let us emphasise nevertheless that the $m$ fixed trace ensembles that we do multiply here are non-Gaussian ensembles.

The partition function for the product matrix $Y_M$ in \eqref{Pidef} is defined as
\bea
\Zd(\{s\}_m;\{t\}_{M-m})&\equiv& \prod_{j=1}^m\int[dG_j]\det[G_jG_j^\dag]^{\nu_j} \delta\left(s_j-\Tr G_jG_j^\dag\right)
\nonumber\\
&&\times\prod_{l=m+1}^M \int[dG_l]\det[G_lG_l^\dag]^{\nu_l}
\exp[-t_l\Tr G_lG_l^\dag], 
\label{ZdMs}
\eea
where $s_1,\ldots,s_m>0$, 
$t_{m+1},\ldots,t_M>0$, and  $\nu_j>-1$ for all $j=1,\ldots,M$. It depends on the set $\{s\}_m$ of all constraint parameters $s_j$ and the set $\{t\}_{M-m}$ of (inverse) variances of the remaining $M-m$ Ginibre ensembles. 
The dependence on all $\nu_j$ is collectively denoted by a subscript $\nu$. Obviously on the level of matrices the partition function \eqref{ZdMs} factorises into $m$  induced Ginibre ensembles with fixed trace constraint, and $M-m$ of such ensembles without constraint. This simply states that all matrices are independent, and as such all parameters $s_j$ and $t_k$ could be scaled out.

For integer $\nu_j$ the setup we consider here is equivalent to the product eq. (\ref{Pidef}) of rectangular matrices with appropriate matrix dimensions, $N_j\times N_{j-1}$ for $j=1,\ldots,M$. Because both the Gaussian Ginibre weight and the weight with a fixed trace constraint are isotropic (or bi-unitarily invariant), we can apply the result from \cite{IK} and reduce the rectangular matrices $N_j\times N_{j-1}$ to square matrices of dimension $N=N_0$, with the induced weights  $\nu_j=N_j-N$ as given above. Furthermore, according to \cite{IK} the order in such a product (\ref{Pidef}) does not matter. The advantage in starting with 
eq. (\ref{ZdMs}) instead, is that we can allow for real parameters $\nu_j>-1$ for all $j=1,\ldots,M$.

In the following Subsection \ref{jpdfY} we will first compute the partition function \eqref{ZdMs} and derive the joint density of complex eigenvalues of the product matrix $Y_M$. Although the latter will turn out to be rather complicated, in the next Subsection \ref{specY} we will give a concise closed form expression for the spectral density of the product matrix $Y_M$ 
for finite $N$. In passing we need to compute the kernel of orthogonal polynomials for the product of $M$ induced Ginibre matrices without constraint, where each matrix has a different variance. This immediately leads to the $k$-point correlation functions for such products.
To access the density correlations functions of $Y_M$, we will use an $m$-fold inverse Laplace transform of such a product of induced Ginibre matrices with different variances, building upon the previous section. For the spectral density this can be made very explicit, and gives also access to the spectrum of the stability exponents in the limit when $M\to\infty$, at least when keeping $m$ much smaller. 
For the $k$-point functions, however,  the nested multiple integral representations become very cumbersome and the $m$-fold inverse Laplace transform cannot be performed. Although the procedure will be formulated they will not be explicitly displayed.

\subsection{Partition function and joint density}\label{jpdfY}

It turns out that a direct calculation of the joint density is not straightforward as for a single fixed trace ensemble in the previous section in \eqref{Zdsev}.  We thus first look at the joint density of the product of only induced Ginibre matrices, with different variances and no constraint, eq. \eqref{Pidef} at $m=0$. Its partition function reads
\be
\Z(\{t\}_M)\equiv \prod_{j=1}^M\int[dG_j]\det[G_jG_j^\dag]^{\nu_j}\exp[-t_j\Tr G_jG_j^\dag] \ ,
\label{ZMdef}
\ee
where $t_1,\ldots,t_M>0$, and the notation is analogous to \eqref{ZdMs}. Again on the level of matrices it factorises into $M$ independent induced Ginibre ensembles.
It is a slight generalisation of \cite{ABu,AIS} in that all (inverse) variances $t_j$ are introduced as additional parameters. 

The following $m$-fold Laplace transform relates the two product ensembles \eqref{ZMdef} and \eqref{ZdMs}, and also fixes our notation,
\be
{\cal L}^m\{\Zd(\{s\}_m;\{t\}_{M-m})\}(\{t\}_m)= \Z
(\{t\}_M)
\ .
\label{ZMtrafo}
\ee
It should be clear that every single Laplace transform ${\cal L}$ from variable $s_j$ to variable $t_j$
acts on one fixed trace ensembles with matrix $G_{j}$, as given in \eqref{Ztrafo}, with $j=1,\ldots,m$.

Before moving to the joint density of \eqref{ZMdef}, we can simply state the parameter dependence of the two partition functions \eqref{ZMdef} and \eqref{ZdMs}, due to factorisation:
\bea
\Z(\{t\}_M) &=& 
(V_N N!)^M \pi^{\frac{MN(N+1)}{2}}\prod_{j=1}^M\frac{1}{ t_j^{N^2+N\nu_j}}
\prod_{l=1}^M\prod_{j=0}^{N-1}\Gamma(j+1+\nu_l)\ ,
\label{ZMtdep}\\
\Zd(\{s\}_m;\{t\}_{M-m}) &=& 
(V_N N!)^M \pi^{\frac{MN(N+1)}{2}}\prod_{j=1}^m\frac{s_j^{N^2+N\nu_j-1}}{\Gamma(N^2+N\nu_j)}\prod_{j=m+1}^M\frac{1}{ t_j^{N^2+N\nu_j}}
\prod_{l=1}^M\prod_{j=0}^{N-1}\Gamma(j+1+\nu_l).
\nonumber\\
\label{ZMsdep}
\eea
The second equation also follows from the first, using the inverse Laplace transform \eqref{Laplace1} $m$ times. 

The joint density of the complex eigenvalues $z_1,\ldots,z_N$ of the product $Y_M=G_1\cdots G_M$ of induced Ginibre matrices without constraint,  including the $t_j$-dependence, follows from a generalised Schur decomposition, $G_j=U_j(Z_j+T_j)U_{j+1}^\dag$, with $U_{M+1}=U_1$ \cite{Eugene,Adhikari}, along the lines of \cite{ABu,AIS}.  Here, the matrices $U_{j=1,\ldots,M}$ are unitary matrices, $T_{j=1,\ldots,M}$ are strictly upper triangular complex matrices, and $Z_j=$diag$(z^{(j)}_1,\ldots,z^{(j)}_N)$ are diagonal complex matrices. They satisfy $Z=Z_1\cdots Z_M$ with $Z=$diag$(z_1,\ldots,z_M)$ containing the complex eigenvalues of matrix $Y_M$. 
In analogy to \eqref{ZGev} we obtain, 
\bea
\Z(\{t\}_M)
&=&
\frac{1}{N!}
\prod_{j=1}^M\frac{V_N\pi^{\frac{N(N-1)}{2}}N!}{t_j^{\frac{N(N-1)}{2}}}
\prod_{a=1}^N\int_{\mathbb C}d^2z_a
\prod_{j=1}^M 
\int_{\mathbb C}d^2z_a^{(j)} |z_a^{(j)}|^{2\nu_j} \e^{-t_j|z_a^{(j)}|^2}
\delta^{(2)}(z_a-z_a^{(1)}\cdots z_a^{(M)})\nn\\
&&
\times |\Delta_N(Z)|^2 
\nonumber\\
&=&\frac{(V_N)^M\pi^{\frac{MN(N-1)}{2}}(N!)^{M-1}}{\prod_{j=1}^Mt_j^{\frac{N(N-1)}{2}}}
\prod_{a=1}^N
\int_{\mathbb C}d^2z_a w(z_a;\{t\}_M)\ |\Delta_N(Z)|^2 
\label{ZMGev}\\
&\equiv& \prod_{a=1}^N
\int_{\mathbb C}d^2z_a P_\nu(z_1,\ldots,z_N;\{t\}_M)\ .
\label{PZMGev}
\eea
In order to see that, we write 
the weight function $w(z_a;\{t\}_M)$ as an $(M-1)$-fold integral, after 
denoting for $|z_a^{(j)}|=r_a^{(j)}$, and 
using the two-dimensional delta-function: 
\bea
w(z_a;\{t\}_M)&=&(2\pi)^{M-1}
\prod_{j=2}^M\int_0^\infty dr_a^{(j)} (r_a^{(j)})^{2\nu_j-1}\e^{-t_j(r_a^{(j)})^2} 
\frac{|z_a|^{2\nu_1}}{(r_a^{(2)}\cdots r_a^{(M)})^{2\nu_1}} 
\e^{-t_1\frac{|z_a|^2}{(r_a^{(2)}\cdots r_a^{(M)})^2}}\nn\\
&=& \pi^{M-1} 
\prod_{j=1}^M t_j^{-\nu_j} 
G^{M\,0}_{0\,M}\left(\mbox{}_{\nu_1,\ldots,\nu_M}^{-} \bigg| \ t_1\cdots t_M|z_a|^2 \right)\ .
\label{w-def}
\eea
Simple substitutions $(q^{(j)})^2=t_j(r^{(j)})^2$ lead to the $(M-1)$-fold integral representation of the Meijer G-function derived in \cite{ABu,AIS}, and we refer to \cite{PeterCoulomb} for a Coulomb gas interpretation of such products of random matrices in \eqref{ZMGev}. 
For the more standard complex contour integral representation of the Meijer G-function we refer to the Appendix, eq. \eqref{Meijer-def}.
Apart from the dependence on the individual  parameters $\nu_j$, the weight function only depends on the product of the variance parameters through 
$t_1\cdots t_M|z|^2$. This will become important later for the $m$-fold inverse Laplace transforms.
As an aside, eqs. \eqref{ZMGev} and \eqref{w-def} constitute the joint density of complex eigenvalues of the product of induced Ginibre matrices with different variances.

In order to derive the joint density of complex eigenvalues for our ensemble \eqref{ZdMs} from \eqref{ZMGev}, we have to perform $m$ inverse Laplace transforms. The difficulty is here that all weight functions in the product over all complex eigenvalues in \eqref{ZMGev} depend on all variances $t_j$, $j=1,\ldots,m$. Therefore, we have to convert back the Meijer G-function in \eqref{w-def} to its $(M-1)$-fold integral representation, in order to be able to collect all dependences on a single $t_j$. 

Let us give an example with $m=1$, before considering the general case.
Collecting the $t_1$-dependence in eqs. (\ref{ZMGev}) and (\ref{w-def}) we can use the inverse Laplace transform eq. (\ref{Laplace2}) to obtain the 
joint density of complex eigenvalues for the product of $m=1$  matrix with a fixed trace constraint with $M-1$ induced Ginibre matrices as
\bea
\Zd(s_1;\{t\}_{M-1})&=&
{\cal L}^{-1}\left\{\Z(\{t\}_M)\right\}(s_1)
\nn\\
&=& \frac{(V_N\pi^{\frac{N(N-1)}{2}}2\pi N!)^{M}}{2\pi N!}
\prod_{a=1}^N\int_{\mathbb C}d^2z_a|z_a|^{2\nu_1}
\prod_{j=2}^M\int_0^\infty dr_a^{(j)}(r_a^{(j)})^{2(\nu_j-\nu_1)-1}\e^{-t_j(r_a^{(j)})^2} \nn\\
&&\times
\frac{\left(s_1-\sum_{a=1}^N\frac{|z_a|^2}{(r_a^{(2)}\cdots r_a^{(M)})^2}\right)^{\frac{N(N-1)}{2}-1}}{\Gamma(\frac{N(N-1)}{2})}
\ |\Delta_N(Z)|^2\ \Theta\left(s_1-\sum_{a=1}^N\frac{|z_a|^2}{(r_a^{(2)}\cdots r_a^{(M)})^2}\right). \nn\\
&\equiv& \prod_{a=1}^N
\int_{\mathbb C}d^2z_a P_{\delta,\nu}(z_1,\ldots,z_N;s_1,\{t\}_{M-1})\ .
\label{PZdm1ev}
\eea
Not only do we loose the determinantal form of the joint density, compared to \eqref{ZMGev}. In contrast to the single fixed trace ensemble \eqref{Zdsev}, the joint density itself contains 
$N(M-1)$ integrals that are nested, making it a highly nontrivial expression. It is clear that further inverse Laplace transform using \eqref{Laplace2} will introduce a further nesting, and we obtain for general $M\geq m\geq 1$
\bea
\Zd(\{s\}_m;\{t\}_{M-m})&=&
{\cal L}^{-m}\left\{\Z(\{t\}_M)\right\}(\{s\}_m)\nn\\
&=& \frac{(V_N)^M \pi^{\frac{MN(N-1)}{2}}(2\pi N!)^{M-1}}{\Gamma\left(\frac{N(N-1)}{2}\right)^m}
\prod_{a=1}^N\int_{\mathbb C}d^2z_a|z_a|^{2\nu_1}
\prod_{j=2}^M\int_0^\infty dr_a^{(j)}(r_a^{(j)})^{2(\nu_j-\nu_1)-1} \nn\\
&\times&
\left[\prod_{j=2}^m \left(s_j-\sum_{a=1}^N(r_a^{(j)})^2\right)\left(s_1-\sum_{a=1}^N\frac{|z_a|^2}{(r_a^{(2)}\cdots r_a^{(M)})^2}\right)\right]^{\frac{N(N-1)}{2}-1}
|\Delta_N(Z)|^2\nn\\
&\times&\prod_{k=m+1}^M\e^{-t_k(r_a^{(k)})^2}
\prod_{j=2}^m \Theta\! \left(s_j-\sum_{a=1}^N(r_a^{(j)})^2\right)
\Theta\!\left(s_1-\sum_{a=1}^N\frac{|z_a|^2}{(r_a^{(2)}\cdots r_a^{(M)})^2}\right)\ \ 
\label{ZMdjpdf}\\
&\equiv& \prod_{a=1}^N
\int_{\mathbb C}d^2z_a P_{\delta,\nu}(z_1,\ldots,z_N;\{s\}_m,\{t\}_{M-m})\ .
\label{PZdmev}
\eea
This exceedingly complicated form for the joint density for general $m$ makes the computation of general $k$-point correlation functions very hard. Also the analysis of the distribution of the radii is difficult as the integration over the angles does not simpify the nensting of the auxiliary integral. Therefore an analysis of the stability exponents bases on the joint density of the radii as it was performed in \cite{ABK} does not seem feasible.  
However, we will see in the next subsection that for general $m$ the spectral density can still be written in a closed, very compact form. 
This fact can the be exploited to directly analyse the density of stability exponents from that.
 
 \subsection{Spectral density and stability exponents}\label{specY}

We start by extending the known results \cite{ABu,AIS} for the product of $M$ induced Ginibre matrices to different variances. To these we can then apply the $m$ inverse Laplace transforms.
The $k$-point density correlation functions are defined as in eq. (\ref{Rkdef}) by integrating out $(N-k)$ complex eigenvalues from the joint density in \eqref{ZMGev}
\bea
R_{\nu}^{(k)}(z_1,\ldots,z_k;\{t\}_M)&\equiv& \frac{N!}{(N-k)!}\frac{1}{\Z(\{t\}_M)}
\frac{(V_NN!\pi^{\frac{N(N-1)}{2}})^M}{N!\prod_{j=1}^Mt_j^{\frac{N(N-1)}{2}}}\nn\\ 
&&\times\int_{\mathbb C}d^2z_{k+1}\ldots\! \int_{\mathbb C}d^2z_N \prod_{a=1}^N w(z_a;\{t\}_M)
|\Delta_N(Z)|^2 .
\label{RkM-def}
\eea
Being a determinantal point process they can be written as the determinant of a kernel as in eq. (\ref{Rkdet}). Because the weight function \eqref{w-def} only depends on the modulus of the argument, its orthogonal polynomials are monic, with the following squared norms:
\bea
\int_{\mathbb C}d^2z\ w(z;\{t\}_M) z^kz^{*\,l}&=&
\delta_{k,l} \frac{\pi^{M-1}}{\prod_{j=1}^Mt_j^{\nu_j}}\ 2\pi\int_0^\infty dr\,r^{2k+1} G^{M\,0}_{0\,M}\left(\mbox{}_{\nu_1,\ldots,\nu_M}^{-} \bigg| \ t_1\cdots t_Mr^2 \right)
\nn\\
&=&
\delta_{k,l} \frac{\pi^{M}}{\prod_{j=1}^Mt_j^{\nu_j+k+1}}\prod_{i=1}^M \Gamma(\nu_i+k+1)\nn\\
&\equiv&
\delta_{k,l} h_k(\{t\}_M)\ .
\label{Mnorms}
\eea
As already used in (\ref{Zprodh}), a consistency check expresses the partition function as the product of these squared norms,
\be
\Z(\{t\}_M)= 
\frac{(V_NN!)^M\pi^{\frac{MN(N-1)}{2}}}{N!\prod_{j=1}^Mt_j^{\frac{N(N-1)}{2}}} \ N! \prod_{k=0}^{N-1} h_k(\{t\}_M)\ ,
\label{ZMprodh}
\ee
which leads back to eq. (\ref{ZMtdep}). 
The kernel of monic orthogonal polynomials resulting from \eqref{Mnorms} is given by 
\bea
K_{\nu}(z,u^*;\{t\}_M)&\equiv&(w(z;\{t\}_M)w(u;\{t\}_M))^{\frac12}\sum_{k=0}^{N-1}\frac{(zu^*)^k}{h_k(\{t\}_M)}\nn\\
&=& G^{M\,0}_{0\,M}\left(\mbox{}_{\nu_1,\ldots,\nu_M}^{-} \bigg| \tau |z|^2 \right)^{\frac12}
G^{M\,0}_{0\,M}\left(\mbox{}_{\nu_1,\ldots,\nu_M}^{-} \bigg| \tau |u|^2 \right)^{\frac12}
\sum_{k=0}^{N-1}\frac{(zu^*)^k \tau^{k+1}}{\pi\prod_{j=1}^M\Gamma(\nu_j+k+1)},\ \ \ \ \ \ 
\label{kernelGM}
\eea
after cancelling some factors. Here, we define $\tau\equiv t_1\cdots t_M$ for the product of all variance parameters. 
The $k$-point correlation functions for the product of induced Ginibre matrices with unequal variances thus read
\be
R_{\nu}^{(k)}(z_1,\ldots,z_k;\{t\}_M)=\det_{1\leq i,j\leq k}\left[ K_{\nu}(z_i,z_j^*;\{t\}_M)\right]\ .
\label{RkMdet}
\ee
This is our first result of this subsection. 
The fact that all correlation functions only depend on the product $\tau$ slightly extends the known results about such products \cite{ABu,AIS,Adhikari}.
For example the spectral density (1-point correlation function) is given by the kernel \eqref{kernelGM} at equal arguments $u=z$, 
\be
R_{\nu}^{(1)}(z_1;\{t\}_M)=K_{\nu}(z,z^*;\{t\}_M)= G^{M\,0}_{0\,M}\left(\mbox{}_{\nu_1,\ldots,\nu_M}^{-} \bigg| \tau |z|^2 \right)
\sum_{k=0}^{N-1}\frac{\tau^{k+1}|z|^{2k}}{\pi\prod_{j=1}^M\Gamma(\nu_j+k+1)}\ .
\label{Gindensity}
\ee
The very simple dependence on the product of all variance parameters $\tau$ implies that the following simple rescaling,
\be
\tau^{-k}R_{\nu}^{(k)}(z_1/\sqrt{\tau},\ldots,z_k/\sqrt{\tau};\{t\}_M) =R_{\nu}^{(k)}(z_1,\ldots,z_k;\{t=1\}_M)\ ,
\label{Rkscale}
\ee
leads to the $k$-point correlation function for the product of induced Ginibre matrices with all variance parameters equal to unity, $t_{j=1,\ldots,M}=1$. This statement  implies the universality of the limiting kernels and correlation functions for our ensemble of induced Ginibre matrices with un-equal variances.

Let us turn to the correlation functions for products of matrices including fixed trace constraints, notably the spectral density. In analogy to \eqref{Rktrafo} and in view of \eqref{ZMtrafo} we can obtain them from an $m$-fold inverse Laplace transform of \eqref{RkMdet}
\be
\Zd(\{s\}_m;\{t\}_{M-m})R_{\delta,\nu}^{(k)}(z_1,\ldots,z_k;\{s\}_m;\{t\}_{M-m})=
{\cal L}^{-m}\left\{\Z(\{t\}_M)R_{\nu}^{(k)}(z_1,\ldots,z_k;\{t\}_M)\right\}(\{s\}_{m}).
\label{RkMtrafo}
\ee
The difficulty is that for general $k$ we have a product of $k$ Meijer G-functions that together with the kernels all depend on all variance parameters. Already a single inverse Laplace transform of the product of $k$ such Meijer G-function does not have a simple expression, e.g. for $k=2$ this leads to a generalised Meier G-function with two arguments, as introduced by \cite{Ag}.
An alternative is to use the $(M-1)$-fold integral representation for each Meijer G-function, in order to take the inverse Laplace transform, as in the derivation of \eqref{ZMdjpdf}. All $k(M-1)$ integrals  then become nested. Obviously this makes the iteration of inverse Laplace transforms cumbersome.

In contrast, for a single Meijer G-function as it appears in the spectral density, the iteration of the inverse Laplace transform remains simple, due to the following general identity for the single inverse Laplace transform of a Meijer G-function:
\be
{\cal L}^{-1}\left\{ \frac{1}{t^b}
G^{m\,n}_{p\,q}\left(\mbox{}_{a_1,\ldots,a_q}^{b_1,\ldots,b_p} \bigg| \ tx \right)
\right\}(s)=s^{b-1}
G^{m\,n}_{p+1\,q}\left(\mbox{}_{a_1,\ldots,a_q}^{b_1,\ldots,b_p,b} \bigg| \ \frac{x}{s} \right)
\label{L-1Gid}
\ee
Because this relation is central to what follows we give a short verification in Appendix \ref{AppMeijer1}.
In our particular case this leads to the following expression for the $m$-fold iteration of inverse Laplace transforms:
\bea
&&
{\cal L}^{-m}\left\{ \frac{1}{\prod_{j=1}^Mt_j^{b_j}}
G^{M\,0}_{0\,M}\left(\mbox{}_{\nu_1,\ldots,\nu_M}^{-} \bigg| \ t_1\cdots t_M |z|^2 \right)
\right\}(s_1,\ldots,s_m)
=\nn\\
&&
=\frac{\prod_{j=1}^m s_j^{b_j-1}}{\prod_{k=m+1}^Mt_k^{b_k}}
G^{M\,0}_{m\,M}\left(\mbox{}_{\nu_1,\ldots,\nu_M}^{b_1,\ldots,b_m} \bigg| \ \frac{t_{m+1}\cdots t_M|z|^2}{s_1\cdots s_m} \right)
\label{L-mGid}
\eea
Let us apply this identity to the spectral density, eq. \eqref{RkMtrafo} for $k=1$:
\bea
&&\Zd(\{s\}_m;\{t\}_{M-m})R_{\delta,\nu}^{(1)}(z_1;\{s\}_m;\{t\}_{M-m})=
{\cal L}^{-m}\left\{\Z(\{t\}_M)R_{\nu}^{(1)}(z_1;\{t\}_M)\right\}(\{s\}_{m})\nn\\
&=&{\cal L}^{-m}\left\{
\frac{(V_N N!)^M \pi^{\frac{MN(N+1)}{2}}\prod_{l=1}^M\prod_{j=0}^{N-1}\Gamma(j+1+\nu_l)}{\prod_{j=1}^M{ t_j^{N^2+N\nu_j}}}
\right.\nn\\
&&\ \ \ \ \ \ \ \ \times\left.
\sum_{k=0}^{N-1}\frac{|z|^{2k} (t_1\cdots t_M)^{k+1}G^{M\,0}_{0\,M}\left(\mbox{}_{\nu_1,\ldots,\nu_M}^{-} \bigg|  t_1\cdots t_M|z|^2 \right)}{\pi\prod_{j=1}^M\Gamma(\nu_j+k+1)}
\right\}(\{s\}_{m})\nn\\
&=&
\frac{(V_N N!)^M \pi^{\frac{MN(N+1)}{2}}\prod_{l=1}^M\prod_{j=0}^{N-1}\Gamma(j+1+\nu_l)}{\prod_{j=m+1}^M{ t_j^{N^2+N\nu_j}}}
\nn\\
&&\times
\sum_{k=0}^{N-1}\frac{|z|^{2k} (t_{m+1}\cdots t_M)^{k+1}
\prod_{j=1}^m s_j^{N^2+N\nu_j-k-2}
G^{M\,0}_{m\,M}\left(\mbox{}_{\nu_1,\ldots,\nu_M}^{N^2+N\nu_1-k-1,\ldots, N^2+N\nu_m-k-1} \bigg|  \frac{t_{m+1}\cdots t_M|z|^2}{s_1\cdots s_m} \right)}{\pi\prod_{j=1}^M\Gamma(\nu_j+k+1)}\!.\nn\\
\label{R1dm}
\eea
After cancelling common factors from \eqref{ZMsdep} on left and right hand side, we obtain the final answer for the spectral density of complex eigenvalues of the product of $m$ fixed trace and $M-m$ induced Ginibre ensembles:
\bea
R_{\delta,\nu}^{(1)}(z_1;\{s\}_m;\{t\}_{M-m})&=&\frac{t_{m+1}\cdots t_M}{s_1\cdots s_m} \prod_{j=1}^m \Gamma(N^2+N\nu_j)\nn\\
&&\times
\sum_{k=0}^{N-1}\frac{G^{M\,0}_{m\,M}\left(\mbox{}_{\nu_1+k,\ldots,\nu_M+k}^{N^2+N\nu_1-1,\ldots, N^2+N\nu_m-1} \bigg|  \frac{t_{m+1}\cdots t_M|z|^2}{s_1\cdots s_m} \right)}{\pi \prod_{j=1}^M\Gamma(\nu_j+k+1)}.
\label{R1dmfinal}
\eea
This is the second and main result of this section. Here, we have used the following property of the Meijer G-function \cite{Grad}
\be
x^k G^{m\,n}_{p\,q}\left(\mbox{}_{a_1,\ldots,a_q}^{b_1,\ldots,b_p} \bigg|  x \right)=
G^{m\,n}_{p\,q}\left(\mbox{}_{a_1+k,\ldots,a_q+k}^{b_1+k,\ldots,b_p+k} \bigg| x \right),
\label{Gmult}
\ee
in multiplying the remaining dependence on $|z|^2$, $t_{j>m}$ and $s_{j\leq m}$ under the sum in \eqref{R1dm} into the Meijer G-function.
Equation \eqref{R1dmfinal} is a remarkably compact expression. Furthermore, the following rescaling of \eqref{R1dmfinal} which is analogous to \eqref{Rkscale} holds, 
\be
\frac{s_1\cdots s_m}{t_{m+1}\cdots t_M}R_{\delta,\nu}^{(1)}\left(\frac{(s_1\cdots s_m)^{\frac12}}{(t_{m+1}\cdots t_M)^{\frac12}}z_1;\{s\}_m;\{t\}_{M-m}\right)
=R_{\delta,\nu}^{(1)}(z_1;\{s=1\}_m;\{t=1\}_{M-m}),
\label{RMmscale}
\ee
making the spectral density independent of all constraint and variance parameters (set to unity here on the right hand side).

For comparison we apply the same identity \eqref{Gmult} to the density \eqref{Gindensity} of the product of $M$ Ginibre matrices 
\be
R_{\nu}^{(1)}(z_1;\{t\}_M)= \tau 
\sum_{k=0}^{N-1}\frac{G^{M\,0}_{0\,M}\left(\mbox{}_{\nu_1+k,\ldots,\nu_M+k}^{-} \bigg| \tau |z|^2 \right)}{\pi\prod_{j=1}^M\Gamma(\nu_j+k+1)}\ .
\label{Gindensity2}
\ee
Apart from the normalisation the only effect of muliplying $m$ FTE is to add a nonzero index to the Meijer G-function under the sum. 
We will see below how the two densities can be mapped in the large-$N$ limit.

\begin{figure}[t]
\centerline{\includegraphics[scale=0.6]{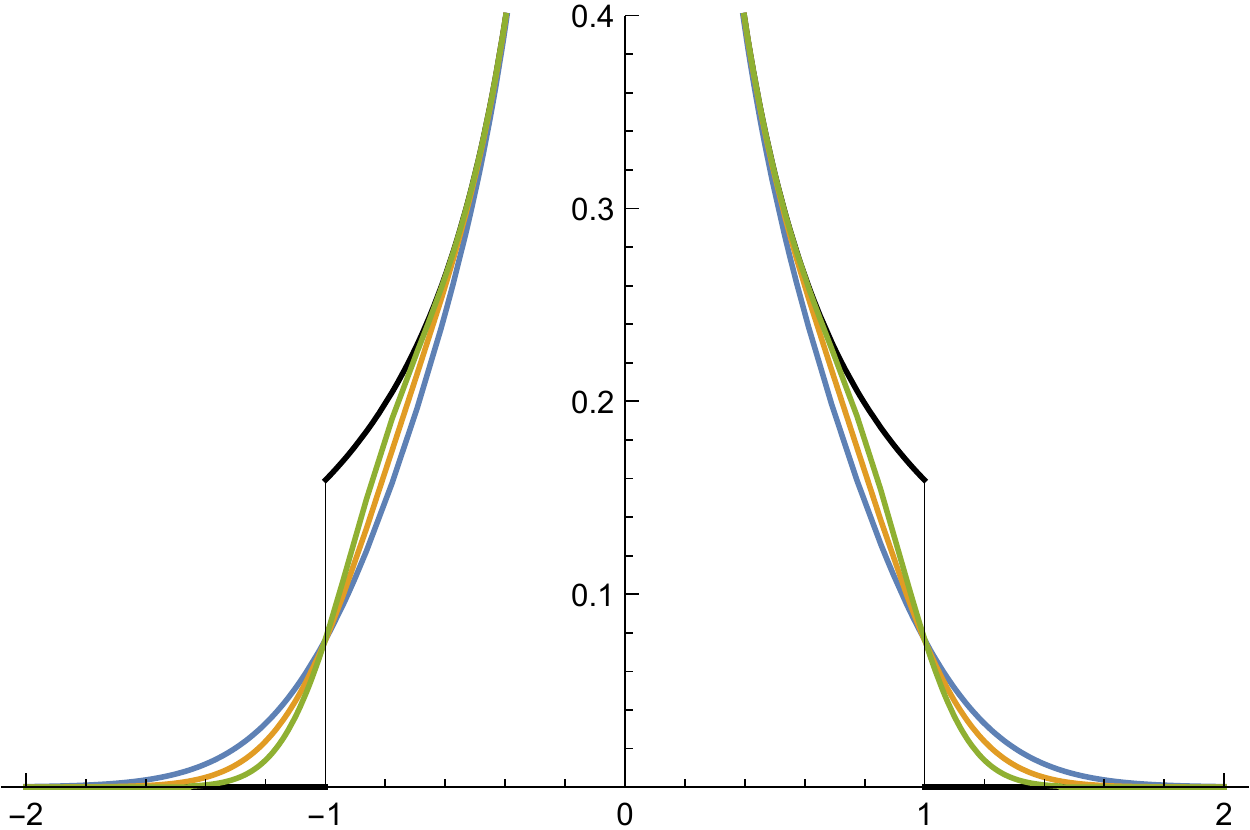}
\includegraphics[scale=0.6]{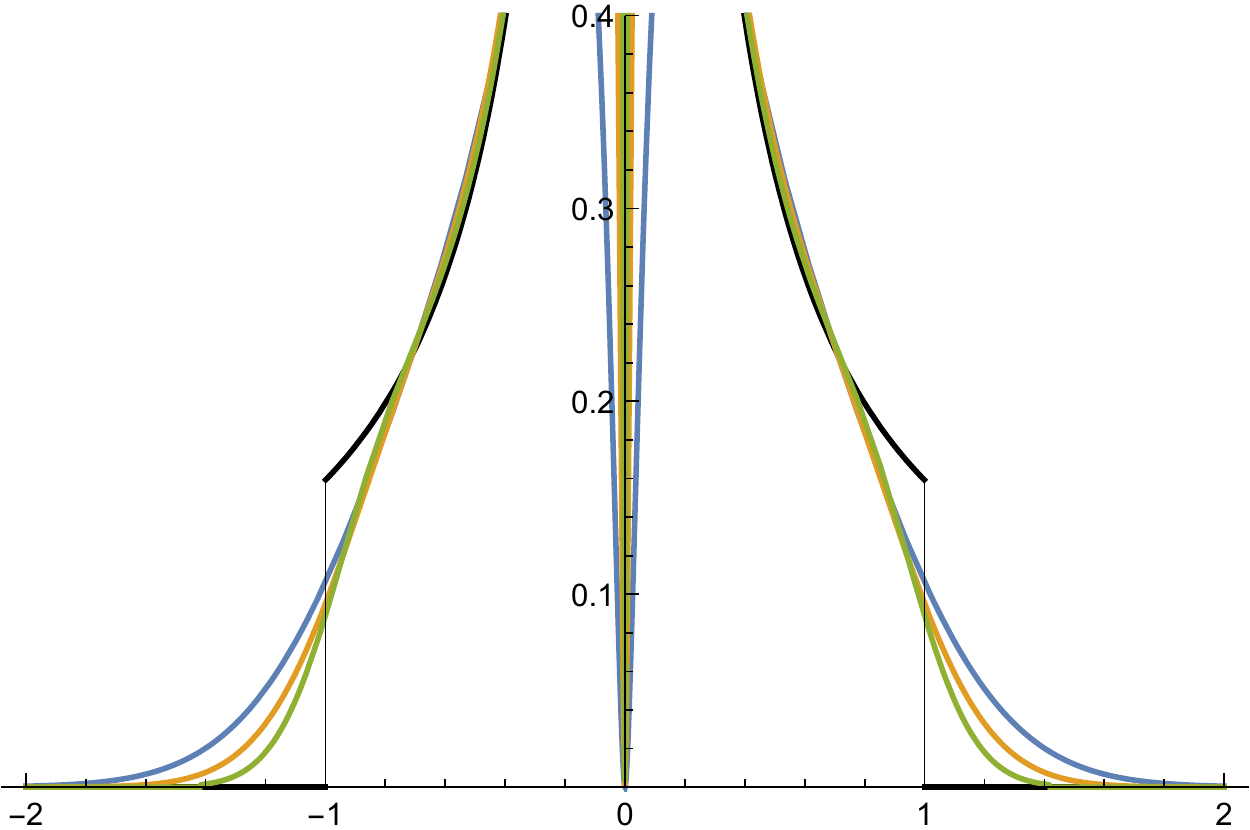}
}
\centerline{\includegraphics[scale=0.6]{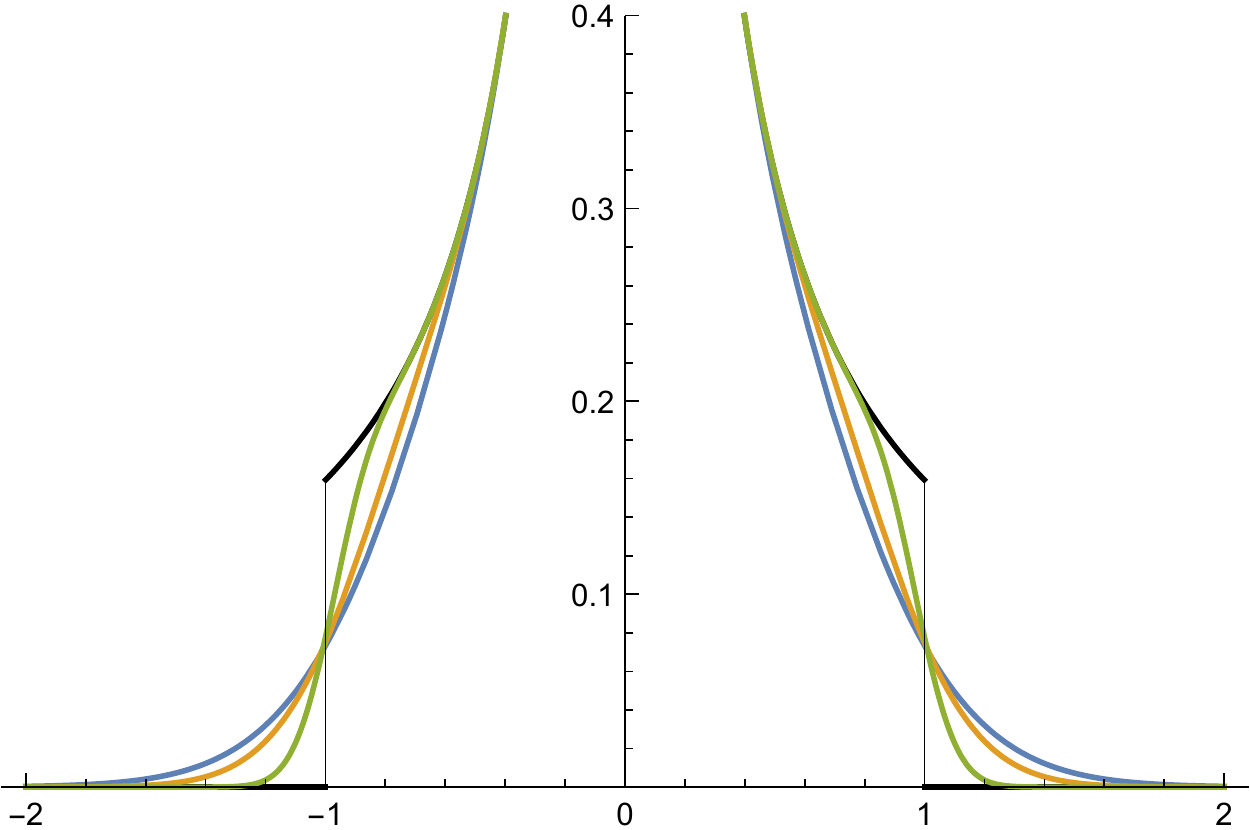}\includegraphics[scale=0.6]{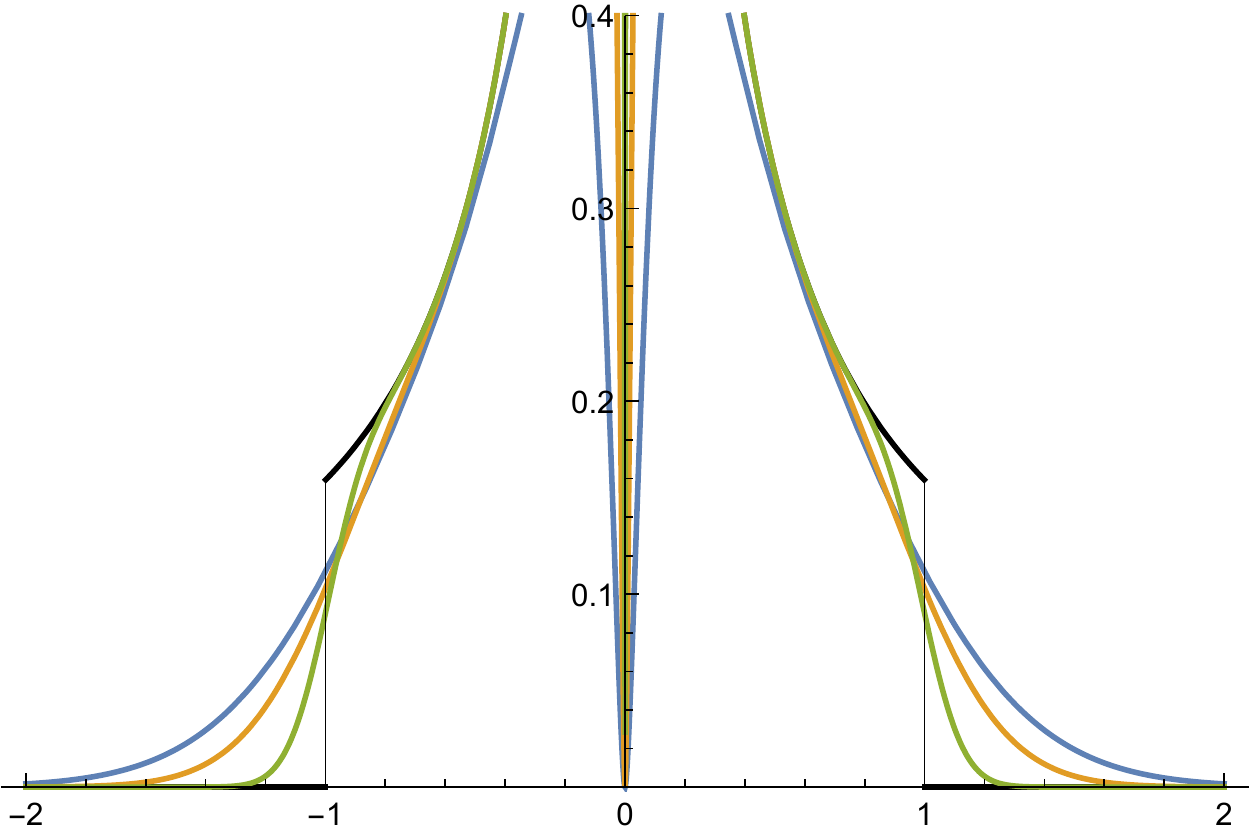}}
\caption{
A cut through the rescaled spectral density 
$R_{\delta,\nu}^{(1)}(z;s_1=1,t_2=1)/N$ is plotted for $z=x+i0$ on $x\in[-2,2]$ for the product of $m=1$ fixed trace with $M-m=1$ induced Ginibre ensemble from \eqref{R1dmfinal} for 
$\nu=0$ (top left) and $\nu=1$ (top right). With moderate numerical effort we can go to values of $N=5,10,20$, corresponding to the bottom blue, middle orange and top green line, respectively. 
For comparison the scaled spectral density $R_{\nu}^{(1)}(zN;t_1=1,t_2=1)N$ of the product of two independent induced Ginibre ensemble from \eqref{kernelGM} is shown for $\nu_1=\nu_2=0$ (bottom left) and $\nu_1=\nu_2=1$ (bottom right). 
Here, we can easily reach values of $N=5,10$, up to $50$,  with the same colour code for increasing $N$. The corresponding limiting global density $\rho(z)=(2\pi|z|)^{-1}\Theta(1-|z|)$ is also given for comparison here and in the above plots (black line).
}
\label{densityM=2Gin-FT}
\end{figure}

In Fig. \ref{densityM=2Gin-FT} we compare an example for the rescaled density of the product of one fixed trace ($m=1$) with one induced Ginibre matrix ($M-m=1$) on the one hand, and the product of $M=2$ independent induced Ginibre matrices on the other hand.
Because the divergence of the limiting global density is so strong, which is known to be
$\sim1/r$ for the product of $M=2$ independent Ginibre matrices
\cite{Burda,Burda2,GG}, the convergence to a compact support 
is less visible than for a single FTE in Fig. \ref{densityM=1Gin-FT}. 
Note again the difference in rescaling for the fixed trace and induced Ginibre ensemble, also compared to $M=1$ in Fig. \ref{densityM=1Gin-FT}, in order to approach the same large-$N$ limit. 

Let us again explain the different scaling used in Fig.  \ref{densityM=2Gin-FT}, now for general $m$ and $M$, which also leads to the universality of the density \eqref{R1dmfinal}.
It is known that the following scaling of the density \eqref{Gindensity} of the product of only independent induced Ginibre matrices ($m=0$) leads to a compact support:
\be
N^{M-1}R_{\nu}^{(1)}\left(zN^{\frac{M}{2}};\{t=1\}_M\right)= N^{M-1}G^{M\,0}_{0\,M}\left(\mbox{}_{\nu_1,\ldots,\nu_M}^{-} \bigg| N^M |z|^2 \right)
\sum_{k=0}^{N-1}\frac{\left(N^M|z|^2\right)^k}{\pi\prod_{j=1}^M\Gamma(\nu_j+k+1)}.
\label{Ginscaled}
\ee
The resulting global limiting density $\rho_M(z)=\frac{1}{M\pi}|z|^{\frac{2}{M}-2}\Theta(1-|z|)$ for $M$ independent matrices 
is known to replace the circular law at $M=1$, c.f. \cite{Burda,Burda2,GG}. 
We now compare with the asymptotic of the density \eqref{R1dmfinal}, where we use the definition \eqref{Meijer-def} for the Meijer G-function, together with \eqref{Gmult} as a starting point:
\bea
R_{\delta,\nu}^{(1)}(z;\{s=1\}_m;\{t=1\}_{M-m})&=& \sum_{k=0}^{N-1}\frac{|z|^{2k}\int_L \frac{dv}{2\pi i}\,|z|^{-2v}\prod_{i=1}^m\frac{\Gamma[N^2+N\nu_i]}{\Gamma[N^2+N\nu_i-k-1+v]}\prod_{l=1}^M\Gamma[\nu_l+v]}{\pi\prod_{j=1}^M\Gamma(\nu_j+k+1)}
\nn\\
&\simeq& \sum_{k=0}^{N-1}\frac{N^{-2m(k+1)}|z|^{2k}\int_L \frac{dv}{2\pi i}\,N^{2mv}|z|^{-2v}\prod_{l=1}^M\Gamma[\nu_l+v]}{\pi\prod_{j=1}^M\Gamma(\nu_j+k+1)}\ ,
\label{limheur}
\eea
for $N\gg1$.
The ratio of Gamma-functions in the numerator cancels to leading order, up to $v$-dependent powers of $N$. This leads to the difference in scaling of $z$. The
remaining complex contour integral in the numerator simply gives $G^{M\,0}_{0\,M}(\mbox{}_{\nu_1,\ldots,\nu_M}^{-} |\ |z|^2/N^{2m})$, as in \eqref{Ginscaled}.
Consequently we have to take the following scaling for the density of the product of $m$ fixed trace and $M-m$ Ginibre matrices:
\be
\lim_{N\gg1} N^{M-1-2m}R_{\delta,\nu}^{(1)}\left(zN^{\frac{M}{2}-m};\{s=1\}_m;\{t=1\}_{M-m}\right)
\simeq
 N^{M-1}R_{\nu}^{(1)}\left(zN^{\frac{M}{2}};\{t=1\}_M\right)\ .
 \label{limdensities}
\ee
This gives the scaling used in the example in Fig. \ref{densityM=2Gin-FT} with $M=2$ and $m=1$. At the same time our heuristic argument yields the universality of the density \eqref{R1dmfinal} in the large-$N$ limit, to be in the same universality class as the product of $M$ unconstrained induced Ginibre ensembles. This extends at least heuristically the rigorous universality proof for a single FTE presented in \cite{ACV}.

In the special case of multiplying $M=2$ independent (induced) Ginibre matrices the corresponding Meijer G-function in the weight function \eqref{w-def} simplifies to an elementary function \cite{Grad}, 
\be
G^{2\,0}_{0\,2}\left(\mbox{}_{\nu_1,\nu_2}^{-} \bigg| x^2\right)=2x^{\nu_1+\nu_2}K_{\nu_1-\nu_2}(2x)\ , 
\ee
the modified Bessel function of second kind. 
It is a natural question if such a simplification also occurs here. Indeed, when multiplying one fixed trace matrix with one induced Ginibre matrix, that is $M=2$ with $m=1$, the corresponding Meijer G-function can be expressed in terms of the confluent hypergeometric function of Kummer type $U$:
\be
G^{2,\,0}_{1,\,2}\left(\mbox{}_{b,0}^{a} \bigg| x\right)=\e^{-x}U(a-b,1-b,x)\ .
\label{GUid}
\ee
Because we did not easily find this relation in the literature and in order to be self-contained, we derive it in Appendix \ref{AppMeijer2}. Using \eqref{Gmult} it is equivalent to 
\be
G^{2,\,0}_{1,\,2}\left(\mbox{}_{b+c,c}^{a+c} \bigg| x\right)=
x^cG^{2,\,0}_{1,\,2}\left(\mbox{}_{b,0}^{a} \bigg| x\right)
=x^c\e^{-x}U(a-b,1-b,x)\ .
\ee
Choosing $a+c=N^2+N\nu_1-1$, $b+c=\nu_1+k$ and $c=\nu_2+k$ in \eqref{R1dmfinal} with $M=2$ and $m=1$, we arrive at
\bea
R_{\delta,\nu}^{(1)}(z;s_1,t_2)&=&\frac{t_2}{s_1}\Gamma(N^2+N\nu_1)\ \e^{-\frac{t_2}{s1}|z|^2}
\sum_{k=0}^{N-1}\frac{1}{\pi\Gamma(k+1+\nu_1)\Gamma(k+1+\nu_2)}\nn\\
&&\times
\left(\frac{t_2|z|^2}{s_1}\right)^{k+\nu_2} 
U\left(N(N+\nu_1)-\nu_1-k-1,1-\nu_1+\nu_2,\frac{t_2|z|^2}{s_1}\right).\ \ \ \ 
\label{RdM=2}
\eea
This is the spectral density of the product of one fixed trace and one induced Ginibre matrix.
The same result can be obtained independently by using the integral representation of the Meijer G-function \eqref{w-def} in \eqref{R1dm}, doing the Laplace transform and the using the standard integral representation of the confluent hypergeometric function $U$.
For this function a more rigorous asymptotic analysis than \eqref{limheur} is available \cite{NIST:2010}, leading to the same answer as in \eqref{limdensities}.

At the end of this section let us turn to the analysis of the stability exponents for the mixed product of $m$ fixed trace and $M-m$ induced Ginibre ensembles. Following \cite{ABK} we redefine the radii of the complex eigenvalues as
\begin{equation}
|z_i|=\exp[M\mu_i]
\label{stabdef}
\end{equation}
while keeping the angles untouched, in order to take the large-$M$ limit. These stability exponents can be used in the same way as the Lyapunov exponents, which are based rather on the singular values, to characterise the asymptotic behaviour of chaotic dynamical systems.

In \cite{ABK} is was shown that in the large-$M$ the limiting exponents $\mu_i$ take deterministic values and have a Gaussian distribution around this radius, with a variance proportional to $1/M$. In \cite{ABK} this argument was base on the analysis of the joint density of the radii, which is rather cumbersome in our case. However,  we can use the following the limit of the Meijer G-function derived in \cite{ABK}
\begin{equation}
\lim_{M\to\infty} \frac{2M}{\Gamma(k+1)^M} G^{M\,0}_{0\,M}\left(\mbox{}_{k+1,\ldots,k+1}^{-} \bigg| \exp[2M\mu] \right)=\delta\left( \mu-\frac{\psi(k+1)}{2}\right),
\label{GM-lim}
\end{equation}
for $k=0,1,\ldots$, with $\psi(x)$ denoting the digamma function,
 and directly use it for the explicit expression for the spectral density. We will first demonstrate this by rederiving the results of \cite{ABK} for the product of $M$ Ginibre ensembles, starting from the density \eqref{Gindensity2}. For simplicity we use the rescaling \eqref{RMmscale} to set all variances $t_j$ and auxiliary parameters $s_j$ to unity. We also set all parameters $\nu_j=0\ ,\ \ \forall j$, for simplicity. The change of variables \eqref{stabdef} maps the density \eqref{Gindensity2} to 
\be
R_{0}^{(1)}(\mu;1)= M\, e^{2M\mu} 
\sum_{k=0}^{N-1}\frac{G^{M\,0}_{0\,M}\left(\mbox{}_{k,\ldots,k}^{-} \bigg| e^{2M\mu} \right)}{\pi\Gamma(k+1)^M}\ .
\label{Gindensitymu}
\ee
Using the shift property \eqref{Gmult}, the exponential can be included into the Meijer G-function, shifting the indices by one.
After integrating out the angle which is trivial, and normalising the remaining density by $1/N$ to one, we can directly apply \eqref{GM-lim} to arrive at the density for the stability exponents for finite-$N$ from \cite{ABK}:
\be
\rho_N(\mu) =\lim_{M\to\infty}\frac{2\pi}{N} R_{0}^{(1)}(\mu;1)=  \frac{1}{N} \sum_{k=0}^{N-1}\delta\left( \mu-\frac{\psi(k+1)}{2}\right).
\label{rhomu}
\ee
We note here that in \cite{ABK} the variances of the limiting Gaussian distributions, that become delta-functions when $M\to\infty$, were also computed explicitly, which we do not display here. 

Let us turn to the spectral density \eqref{R1dmfinal} of the mixed product. After the rescaling, setting all $\nu_j=0$ and the change of variables \eqref{stabdef} we obtain
\begin{equation}
R_{\delta,0}^{(1)}(\mu;\{1\}_m;\{1\}_{M-m})=M\Gamma(N^2)^m
\sum_{k=0}^{N-1}\frac{G^{M\,0}_{m\,M}\left(\mbox{}_{k+1,\ldots,k+1}^{N^2,\ldots, N^2} \bigg|  e^{2M\mu}\right)}{\pi \Gamma(k+1)^M},
\label{RMNmu}
\end{equation}
where we have already included the exponential into the Meijer G-function. 
Let us look at the ratio of Gamma functions times the Meijer G-function, using the integral representation \eqref{Meijer-def}. For large $N\gg1$ we can approximate 
\bea
&&\frac{1}{2\pi i}\int_L du\,e^{2M\mu u} \frac{\Gamma(k+1-u)^M}{\Gamma(N^2-u)^m} \frac{\Gamma(N^2)^m}{\Gamma(k+1)^M}\nn\\
\stackrel{N\gg1}{\approx} &&
\frac{1}{2\pi i}\int_L du\,e^{2M\mu u} e^{m\log[N^2]u}\frac{\Gamma(k+1-u)^M}{\Gamma(k+1)^M} \ ,
\label{Gapprox}
\eea
by replacing the ratio $\Gamma(N^2)/\Gamma(N^2-u)\approx (N^2)^u$. If we then also take $M\to\infty$, such that the condition $M\gg m\log[N]$ is satisfied, the $m$-dependence becomes subleading and we arrive at the same results as for the product of $M$ Ginibre matrices \eqref{rhomu},
\be
\lim_{M\to\infty,N\gg1}\frac{2\pi}{N} R_{\delta,0}^{(1)}(\mu;\{1\}_m;\{1\}_{M-m})=\frac{1}{N} \sum_{k=0}^{N-1}\delta\left( \mu-\frac{\psi(k+1)}{2}\right),
\label{rhomuFT}
\ee
which agrees with \eqref{rhomu} and is thus universal. We expect that to next order in $1/M$ also the Gaussian shape and the variances obtained in \cite{ABK} will not change in this limit. The analysis of more general  products, e.g. of only FTE with $m=M$, is more involved and left for future work.

\sect{Conclusions}
\label{conclusio}

In this paper we have looked at the complex eigenvalue correlation functions of complex non-Hermitian matrices, that are subject to a fixed trace constraint on their second moment. There are several physics applications from Coulomb gases, open bipartite quantum systems or chaotic dynamics to motivate such a setup. 
In the first part we have looked at a single of such fixed trace ensembles (FTE). We have compared an ensemble originating from an induced complex Ginibre matrix $G$, 
that is equivalent to a standard Ginibre ensembles with rectangular $G$, to that of induced complex normal matrices. We were motivated by the question whether or not these will constitute different two-dimensional Coulomb gases. As our first result we found that the former FTE of induced Ginibre ensembles leads to a Coulomb gas with an effective potential, subject to a bound on the sum of the squared absolute values of the complex eigenvalues. In that it resembles more a restricted trace ensemble. 
The modification compared to the constraint on matrix space came from the additional degrees of freedom in the Schur decomposition of $G$. For the normal FTE there were no such degrees of freedom and the hard constraint remained. 
Using an inverse Laplace transform of the unconstrained ensembles, our second result revealed that despite these differences, already at finite matrix size $N$ all $k$-point correlation of these two FTE are of the same functional form and thus agree up to a rescaling of $N$. For Hermitian FTE such an agreement between correlation functions of restriced and FTE does not hold.

In the large-$N$ limit we only gave heuristic arguments for the density to agree with that of the induced Ginibre ensemble, as this was shown rigorously elsewhere with our coauthor for a more general FTE. In passing we also obtained an interesting mathematical result, that at the origin the rate of convergence to the circular law in FTE is only algebraic, compared to the known exponential rate for the Ginibre ensemble. Our final result for a single FTE was the computation of the gap probability at the origin, given in terms of an inverse Laplace transform of the corresponding quantity in the Ginibre ensemble. Upon differentiation this gives rise to the distribution of the smallest eigenvalue in radius, which could be relevant in questions of entanglement in open bi-partite quantum systems. We conjecture it to be universal, too.

The second part of this paper was devoted to the complex eigenvalue spectrum of products of $m$ FTE with $M-m$ induced Ginibre ensembles, building upon what we had learned in the first part. Because multiplying normal matrices without constraint is already a difficult open problem we did not consider such matrices here.
Our motivation was the application of products of random matrices to chaotic dynamical systems. In particular we wanted to know, if the recently obtained results for the stability exponents of products of independent induced Ginibre matrices extend to non-Gaussian distributions and are thus universal. For the origin of the spectrum and for the stability exponents, that characterise such dynamical systems, no previous universality results were known.
Previous results for the Lyapunov exponents had shown, that introducing a non-trivial covariance matrix in the Ginibre ensembles does indeed change the locations of the limiting Lyapunov exponents.

Our main tool was again an inverse Laplace transform, which had to be iterated $m$ times here. For our point of departure we first showed, that when multiplying only induced Ginibre matrices with different variances, the resulting correlation functions merely have to be rescaled by the product of all variances.  Next, we derived the joint density of complex eigenvalues for mixed products of $m$ FTE with $M-m$ induced Ginibre ensembles, being given by nested integrals. For that reason we then focussed on the spectral density alone. Surprisingly, the $m$-fold Laplace inversion could be explicitly performed, and we arrived at a closed formula for the density for finite $M$, $m$, $N$ and rectangularity parameters. We exploited this main result of this part furthermore, to give again heuristic arguments for the universality of the spectral density for large-$N$ and fixed $m$ and $M$. Last but not least we addressed the question of stability exponents in this setup of mixed products.
In taking the large $M$-limit together with large $N$,  such that $M\gg m \log[N]$, we could show, that the density of stability exponents is universal here and agrees with that of multiplying independent Ginibre ensembles only. 
We expect that our universality argument can be made rigorous in all different large-$N$ limits, and that it can in particular be extended to all local $k$-point correlations functions, as this was shown for a single FTE with our coworker recently.

\ \\
\noindent
{\bf Acknowledgments:}
\noindent
Support of the German research council DFG through grant AK35/2-1 "Products of Random Matrices"  
and CRC 1283 "Taming uncertainty and profiting from randomness and low regularity in analysis, stochastics and their applications"
(G.A.), as well as through the International Graduate College ``Stochastic and Real World Models'' Beijing-Bielefeld  (M.C.) is gratefully acknowledged. The Simons Center for Geometry and Physics, Stony Brook University is thanked (G.A.) for support and kind hospitality where part of this work was being done. We would like to thank Martin Venker for useful discussions and Mario Kieburg for a careful reading of the manuscript.

\begin{appendix}

\sect{Identities for the density of fixed trace ensembles}\label{AppA}

\subsection{Alternative derivation for the density of the fixed trace Ginibre ensemble}
\label{alternative}

In this part of the appendix we present an alternative derivation of the density of the fixed trace induced Ginibre ensemble eq. (\ref{Rdfinal}), following the ideas of \cite{ACMV1}. 
Our starting point is that the density of the induced Ginibre ensemble \eqref{ZGev} can be computed using polar coordinates and compared to the result from orthogonal polynomials.
The angular integral that we encounter on the way will then be used to compute the density in the ensemble with a fixed trace constraint. 

From the definition \eqref{Rkdef} we have for the density of the induced Ginibre ensemble
\bea
R_{\nu}^{(1)}(z_1;t) &=&\frac{N!}{(N-1)!}\frac{1}{\Z(t)}
\int_{\mathbb C}d^2z_{2}\ldots \int_{\mathbb C}d^2z_N 
\int[dU] \int[dT] \prod_{j=1}^N|z_j|^{2\nu}
\ |\Delta_N(Z)|^2\ \e^{-t\Tr ZZ^*-t\Tr TT^\dag}\nn\\
&=& \frac{NV_N\left(\frac{\pi}{t}\right)^{\frac{N(N-1)}{2}}}{\Z(t)}|z_1|^{2\nu}
\e^{-t|z_1|^2}
 \prod_{j=2}^N\int_{\mathbb C}d^2z_{j}|z_j|^{2\nu}|z_1-z_j|^2\e^{-t|z_j|^2} \ |\Delta_{N-1}(Z^\prime)|^2\nn\\
&=& \frac{NV_N\left(\frac{\pi}{t}\right)^{\frac{N(N-1)}{2}}}{\Z(t)}|z_1|^{2\nu}
\e^{-t|z_1|^2}\int_0^\infty dRR^{N^2+N(1+2\nu)-3-2\nu}\e^{-tR^2}
\Psi(z_1,R)\ ,
\label{psiint}
\eea
with 
\bea
\Psi(z_1,R)&\equiv&\int d\Omega_{2N-2}\prod_{j=2}^N|\phi_j|^{2\nu}\left|\frac{z_1}{R}-\phi_j \right|^2
\ |\Delta(\phi_2,\ldots,\phi_N)|^2
=\sum_{k=0}^{N-1}c_{2k}\frac{|z_1|^{2k}}{R^{2k}}\ .
\label{psidef}
\eea
In the first step we have used the property of the Vandermonde determinant (\ref{Vandermonde}), denoting by $Z^\prime=\mbox{diag}(z_2,\ldots,z_N)$.  
In the second step we have changed to polar coordinates of the vector of real dimension $2N-2$ with squared length $R^2=\sum_{j=2}^N|z_j|^2$. The complex eigenvalues can be parametrised in terms of the angles of the corresponding $(2N-N)$-sphere $\Omega_{2N-2}$ as $z_j=R\phi_j$, where the functions $\phi_j$ depend on these angles. Finally, we have used the homogeneity of all factors including the Vandermonde determinant. 
The remaining group integral $\Psi(z_1,R)$, and in particular all its non-vanishing coefficients $c_{2k}$, can be determined by comparing to the result for the density obtained using orthogonal polynomials, eq. (\ref{Gdensity}), 
\bea
R_{\nu}^{(1)}(z;t)=|z|^{2\nu}\e^{-t|z|^2}\sum_{l=0}^{N-1}
\frac{|z|^{2l}}{\pi\Gamma(l+1+\nu)}t^{l+1+\nu}\ ,\nn
\eea
after doing the radial integral over $R$, when inserting \eqref{psidef} into \eqref{psiint}.
Inserting all factors we obtain
\be
c_{2k}=\frac{2(N-1)!\ \pi^{N-1}\prod_{j=0}^{N-1}\Gamma(j+1+\nu)}{\Gamma(N(N+1)/2+N\nu-\nu-k-1)\Gamma(k+1+\nu)}\ .
\label{c2k}
\ee
The density for the corresponding fixed trace ensemble \eqref{Zdsev}  now follows in a similar fashion. From the definition eq. (\ref{Rkdef}) we have 
\bea
R_{\delta,\nu}^{(1)}(z_1;s) 
&=& \frac{1}{(N-1)!}\frac{\Gamma(N^2+N\nu)}{\pi^N\Gamma\left(\frac{N(N-1)}{2}\right)}
\frac{s^{-N^2-N\nu+1}}{\prod_{j=0}^{N-1}\Gamma(j+1+\nu)}|z_1|^{2\nu}
\prod_{j=1}^N\int_{\mathbb C}d^2z_{j}|z_j|^{2\nu}|z_1-z_j|^2\nn\\
&&\times 
|\Delta_{N-1}(Z^\prime)|^2
\left(s- |z_1|^2-\sum_{l=2}^N|z_l|^2\right)^{\frac{N(N-1)}{2}-1}
\Theta\left(s-|z_1|^2-\sum_{l=2}^N|z_l|^2\right)\nn\\
&=& \frac{1}{(N-1)!}\frac{\Gamma(N^2+N\nu)}{\pi^N\Gamma\left(\frac{N(N-1)}{2}\right)}
\frac{s^{-N^2-N\nu+1}|z_1|^{2\nu}}{\prod_{j=0}^{N-1}\Gamma(j+1+\nu)}
\int_0^\infty dRR^{N^2+N+2\nu(N-1)-3-2\nu}\nn\\
&&\times \Psi(z_1,R) \left(s- |z_1|^2-R^2\right)^{\frac{N(N-1)}{2}-1}
\Theta\left(s-|z_1|^2-R^2\right)\nn\\
&=& \frac{\Gamma(N^2+N\nu)s^{-N^2-N\nu+1}|z_1|^{2\nu}}{(N-1)!\pi^N\prod_{j=0}^{N-1}\Gamma(j+1+\nu)}\sum_{k=0}^{N-1}c_{2k}|z_1|^{2k}
\frac{\Gamma(N(N+1)/2+N\nu-\nu-k-1)}{\Gamma(N^2+N\nu-\nu-k-1)}\nn\\
&&\times(s-|z_1|^2)^{N^2+N\nu-\nu-k-2}\Theta(s-|z_1|^2)\ .
\label{Rdpolar}
\eea
Here, we encounter the same angular integral  as in (\ref{psidef}). Using its coefficients explicitly given in \eqref{c2k}, we can perform the radial integration which is different from the Gaussian case. Upon insertion of eq. (\ref{c2k}) in eq. (\ref{Rdpolar}) we find back eq. (\ref{Rdfinal}) that was derived through Laplace transformation. The same computation can be done for the normal ensemble \eqref{Zdnsev}, see \cite{Milan}.

The approach presented here has the disadvantage that it is not easily generalised to arbitrary $k$-point density correlation functions.

\subsection{Equivalence of Eqs. (\ref{Rdfinal}) and (\ref{Rdfinal-alt}) for the fixed trace Ginibre ensemble}

In this subsection we derive the equivalence between the result for the spectral density (\ref{Rdfinal}) in the induced Ginibre ensemble with a fixed trace constraint, as derived in the main text, and the alternative representation as a power series in a single variable  (\ref{Rdfinal-alt}). 
Let us denote by $r=|z|$ and assume that $r^2\in(0,s)$, dropping the Heaviside-function. 
We begin with \eqref{Rdfinal}, pulling out powers of $(s-r^2)$ and expanding the remaining $l$-dependent power $N-l-\nu-1$
\bea
R_{\delta,\nu}^{(1)}(z;s)
&=&\frac{(N^2+N\nu-1)}{\pi s^{N^2+N\nu-1}}(s-r^2)^{N^2+N\nu-N-1}
\sum_{l=0}^{N-1}
{{N^2+N\nu-2}\choose{l+\nu}}
r^{2l+2\nu}\nn\\
&&\times\sum_{b=0}^{N-l-\nu-1}s^{N-l-\nu-1-b}(-r^2)^b{{N-l-\nu-1}\choose{b}}.
\eea
Changing summation variables in the second sum from $b$ to $m=b+l+\nu$, we can interchange the two sums as follows,
\be
\sum_{l=0}^{N-1}\sum_{m=l+\nu}^{N-1} = \sum_{m=\nu}^{N-1}\sum_{l=0}^{m-\nu}\ ,
\ee
to arrive at 
\bea
R_{\delta,\nu}^{(1)}(z;s)
&=&\frac{(N^2+N\nu-1)}{\pi s}\left(1-\frac{r^2}{s}\right)^{N^2+N\nu-N-1}
\nn\\
&&\times\sum_{m=\nu}^{N-1}\frac{r^{2m}}{s^m}
\sum_{l=0}^{m-\nu}(-1)^{m-l-\nu}{{N^2+N\nu-2}\choose{l+\nu}}{{N-l-\nu-1}\choose{m-l-\nu}}\ .
\eea
The second sum in the last line can be rewritten and after a change of variables from $m$ to $m^\prime=m-\nu$ we arrive at \eqref{Rdfinal-alt}. Only in the case $\nu=0$ the coefficients of the power series in the last line can be simplified, as given in the main text.

\subsection{Inverse Laplace transform for the fixed trace normal ensemble}\label{FTn}

In this subsection we provide a few intermediate steps that lead to the density of the induced normal ensemble with a fixed trace constraint eq. (\ref{Rdnfinal}),
and likewise to the corresponding $k$-pont densities. In analogy to eq. (\ref{RdLaplace})
we have 
\bea
\Zdn(s)R_{\delta,\nu}^{(1)\cal N}(z;s)&=&
{\cal L}^{-1}\left\{\Zn(t)R_{\nu}^{(1)\cal N}(z;t)\right\}(s)\nn\\
&=& {\cal L}^{-1}\left\{
V_N \frac{\pi^N}{t^{N(N+1)/2+N\nu}}
N!\prod_{j=0}^{N-1}\Gamma(j+1+\nu)
\sum_{l=0}^{N-1}
\frac{|z|^{2l+2\nu}\e^{-t|z|^2}}{\pi\Gamma(l+1+\nu)}t^{l+1+\nu}
\right\}(s)\nn\\
&=&
V_N \pi^N N!\prod_{j=0}^{N-1}\Gamma(j+1+\nu)
\sum_{l=0}^{N-1}\frac{|z|^{2l+2\nu}(s-|z|^2)^{\frac{N(N+1)}{2}+N\nu-l-\nu-2}\Theta(s-|z|^2)}{\pi\Gamma(l+1+\nu)
\Gamma(\frac{N(N+1)}{2}+N\nu-l-\nu-1)}.\nn\\
\label{A8}
\eea
The only difference is resulting from the different normalising partition functions \eqref{Zdnt} and \eqref{Znt}, compared to  (\ref{RdLaplace}).
For our parameter values here, 
$$N(N+1)/2+N\nu-l-\nu-1\geq N(N+1)/2+N\nu-N-\nu=(N/2+\nu)(N-1)>0$$ 
is positive for $N>1$. Dividing \eqref{A8} by (\ref{Zdnt}) we arrive at eq. (\ref{Rdnfinal}), the density for the induced normal ensemble with a fixed trace constraint. It differs by the corresponding Ginibre ensemble by the replacement of $N^2\to N(N+1)/2$.

The computation for the $k$-point densities can be achieved in the same way, by adopting the lines in eq. (\ref{RkLaplace}) to the normal ensemble, for $k\leq N-1$. This leads again to the replacement 
$N^2\to\frac{N(N+1)}{2}$ in eq. (\ref{Rkdfinal}), with the final result for $k$-point densities of the fixed trace induced Gaussian normal ensemble reading
\bea
R_{\delta,\nu}^{(k)\, \mathcal{N}}(z_1,\ldots,z_k;s)&=&\frac{\Gamma(\frac{N(N+1)}{2}+N\nu)\prod_{j=1}^k|z_j|^{2\nu}}{s^{\frac{N(N+1)}{2}+N\nu-1}}
\left(s-\sum_{i=1}^k|z_i|^2\right)^{\frac{N(N+1)}{2}+N\nu-1}\Theta\left(s-\sum_{i=1}^k|z_i|^2\right)
\nn\\
&&\times\sum_{\sigma\in S_k}(-1)^\sigma\prod_{j=1}^k
\left(\sum_{l_j=0}^{N-1}\frac{(z_jz_{\sigma(j)}^*)^{l_j}}{\pi\Gamma(l_j+1+\nu)\left(s-\sum_{i=1}^k|z_i|^2\right)^{l_j+1+\nu}}
\right)\nn\\
&&\times \frac{1}{\Gamma\left(\frac{N(N+1)}{2}+N\nu-\sum_{n=1}^kl_n-k(\nu+1)\right)} .
\label{RkdNfinal}
\eea

\sect{Identities for Meijer G-functions}
\label{AppMeijer}

In this appendix we verify the inverse Laplace transform given in \eqref{L-1Gid} as well as the relation to the confluent hypergeometric function in \eqref{GUid}.

\subsection{Inverse Laplace transform}\label{AppMeijer1}

The inverse Laplace transform of the Meijer G-function given in \eqref{L-1Gid} can be easily derived by Laplace transforming the right hand side, using elementary integrals and properties of the Meijer G-function.
We begin with the definition of the Laplace transform of the Meijer G-function, 
\bea
{\cal L}\left\{ s^{b-1} G^{m,\,n}_{p+1,\,q}\left(\mbox{}_{a_1,\ldots,a_q}^{b_1,\ldots,b_p,b} \bigg| \frac{x}{s}\right)\right\}(t)&=&
\int_0^\infty ds\ \e^{-st} s^{b-1} G^{m,\,n}_{p+1,\,q}\left(\mbox{}_{a_1,\ldots,a_q}^{b_1,\ldots,b_p,b} \bigg| \frac{x}{s}\right)\nn\\
&=&\frac{1}{t^b}G^{m+1,\,n}_{p+1,\,q+1}\left(\mbox{}_{b,a_1,\ldots,a_q}^{b_1,\ldots,b_p,b} \bigg| tx\right)\nonumber\\
&=&\frac{1}{t^b}G^{m,\,n}_{p,\,q}\left(\mbox{}_{a_1,\ldots,a_q}^{b_1,\ldots,b_p} \bigg| tx\right).
\label{L-1proof}
\eea
In the first step we have used an elementary integral of the Meijer G-function, see e.g. eq. (A.5) in \cite{AIK}, after substituting $st=r$. 
The index $b$ appears twice, and cancels from the definition \eqref{Meijer-def} given below.
This leads to the desired identity, the untransformed left hand side of \eqref{L-1Gid}.
An alternative derivation directly uses the complex contour integral representations \eqref{Meijer-def}  and applies the inverse Laplace transform to it.

\subsection{Relation to the confluent hypergeometric function}\label{AppMeijer2}

We derive an identity relating the Meijer G-function appearing in the product of $M=2$ random matrices, one induced Ginibre and one fixed trace, to the confluent hypergeometric function of Kummer type $U$.
The standard definition of the Meijer G-function in terms a complex contour integral is given by \cite{NIST:2010}
\be
G^{m,\,n}_{p,\,q}\left(\mbox{}_{b_1,\ldots,b_q}^{a_1,\ldots,a_p} \bigg| x\right)
\equiv 
\frac{1}{2\pi i}\int_L du\,x^u\frac{\prod_{i=1}^m\Gamma[b_i-u]\prod_{i=1}^n\Gamma[1-a_i+u]}{\prod_{i=n+1}^p\Gamma[a_i-u]\prod_{i=m+1}^q\Gamma[1-b_i+u]}\ .
\label{Meijer-def}
\ee
The integration contour $L$ depends on the location of the poles of the Gamma-functions, and we refer to \cite{NIST:2010} for the different possibilities. 
In our particular case we obtain the following representation, after the change of variables $u=-t$:
\be
G^{2,\,0}_{1,\,2}\left(\mbox{}_{b,0}^{a} \bigg| x\right)=\frac{1}{2\pi i}\int_{-i\infty}^{+i\infty} dt\,x^{-t}\frac{\Gamma[b+t]\Gamma[t]}{\Gamma[a+t]}\ .
\label{G2012}
\ee
Let us repeat the identity we wish to show,
\be
G^{2,\,0}_{1,\,2}\left(\mbox{}_{b,0}^{a} \bigg| x\right)=\e^{-x}U(a-b,1-b,x)\ .
\label{Bid}
\ee
Using the connection formula \cite[Eq.13.2.40]{NIST:2010} for the confluent hypergeometric function $U$, we have
\be
U(a-b,1-b,x)=x^b U(a,1+b,x)\ .
\ee
In this form we can apply the Mellin-Barnes complex contour integral representation of $U$, \cite[Eq.13.4.18]{NIST:2010}, to the exponential function times $U$ on the right-hand side of \eqref{Bid}:
\be
\e^{-x}x^b U(a,1+b,x)= x^b x^{1-(1+b)}\frac{1}{2\pi i}\int_{-i\infty}^{+i\infty} dt\,x^{-t}\frac{\Gamma[(1+b)-1+t]\Gamma[t]}{\Gamma[a+t]}\ ,
\ee
which agrees with the complex contour integral representation \eqref{Meijer-def} of the Meijer G-function in \eqref{G2012} on the left hand side of the identity.

\end{appendix}



\begin{thebibliography}{99}

\bibitem{Adhikari}  K. Adhikari, N.K. Reddy, T.R. Reddy, and K. Saha,  
Ann. Inst. H. Poincar\'e Probab. Statist. Volume {\bf 52}, Number 1 (2016), 16-46
[arXiv:1308.6817 [math.PR]].

\bibitem{Ag} R.P. Agarwal,
 Proc. Nat. Inst. Sci. India, Vol. {\bf 31 A}, No. 6 (1965), 537-546.

\bibitem{A01} G. Akemann, Phys. Rev. {\bf D64} (2001) 114021 [hep-th/0106053]. 

\bibitem{ACMV1}
G. Akemann, G.M. Cicuta, L. Molinari, and G. Vernizzi,
Phys. Rev. {\bf E59} (1999) 1489-1497
[arXiv:cond-mat/9809270].

\bibitem{ACMV2} G. Akemann, G.M. Cicuta, L. Molinari, and G. Vernizzi,
Phys. Rev. {\bf E60} (1999) 5287 [arXiv:cond-mat/9904446].



\bibitem{AV}
  G. Akemann and G. Vernizzi,
  Nucl.\ Phys.\  {\bf B583} (2000) 739-757
  [arXiv:hep-th/0002148].

\bibitem{APS} G. Akemann, M.J. Phillips, L. Shifrin,
  J. Math. Phys. {\bf 50} (2009)  063504 
[arXiv:0901.0897 [math-ph]].

\bibitem{ABu} G. Akemann and Z. Burda, J. Phys. A: Math. Theor. {\bf 45} (2012)  465201 [arXiv:1208.0187 [math-ph]].

\bibitem{AKW} G. Akemann, M. Kieburg, and L. Wei,
J. Phys. A: Math. Theor. {\bf 46} (2013) 275205, 24pp
[arXiv:1303.5694 [math-ph]].

\bibitem{AIK} G. Akemann, J. Ipsen, and M. Kieburg, Phys. Rev. {\bf E88} (2013) 052118 [arXiv:1307.7560 [math-ph]].

\bibitem{ABKN} G. Akemann, Z. Burda, M. Kieburg, and T. Nagao, 
J. Phys. A: Math. Theor. {\bf 47} (2014) 255202
[arXiv:1310.6395].

\bibitem{AIS} G. Akemann, J.R. Ipsen, and E. Strahov,
Random Matrices: Th. Appl. {\bf 3}, no. 4 (2014) 1450014
[arXiv:1404.4583v1 [math-ph]].

\bibitem{ABK}  G. Akemann, Z. Burda, and M. Kieburg,
J. Phys. A: Math. Theor. {\bf 47} (2014) 395202 
[arXiv:1406.0803 [math-ph]].

\bibitem{AIp}
G. Akemann and J.R.  Ipsen, 
Acta Phys. Polon. B, Vol. {\bf 46}, no. 9 (2015) 1747--1784.
[arXiv:1502.01667 [math- ph]].

\bibitem{ACV} G. Akemann, M. Cikovic, and M. Venker,
Universality at weak and strong non-Hermiticity beyond the elliptic Ginibre ensemble,
arXiv:1610.06517v2 [math.PR].

\bibitem{ameur}
Y. Ameur, H. Hedenmalm, and N. Makarov,
Duke Math. J.
Volume {\bf 159}, Number 1 (2011), 31-81.
[arXiv:0807.0375v3 [math.PR]].

\bibitem{berman}
R.J. Berman,
Determinantal point processes and fermions on complex manifolds: Bulk
universality,
arXiv:0811.3341v1 [math.CV].

\bibitem{JPB} J.-P. Blaizot, J. Grela, M. A. Nowak, W. Tarnowski, P. Warchol
J. Stat. Mech. (2016) 054037
[arXiv:1512.06599].

\bibitem{BD}
P. Bourgade and G. Dubach, The distribution of overlaps between eigenvectors of Ginibre matrices, arXiv:1801.01219 [math.PR].


\bibitem{Burda}
Z. Burda, R. A. Janik, and B. Waclaw,
Phys. Rev. {\bf E81} (2010) 041132 [arXiv:0912.3422v2 [cond-mat.stat-mech]].

\bibitem{Burda2}
Z. Burda, A. Jarosz, G. Livan, M. A. Nowak, and A. Swiech,
Phys. Rev. {\bf E82} (2010) 061114
[arXiv:1007.3594v1 [cond-mat.stat-mech]].

\bibitem{Chen}
Y. Chen, D.-Z. Liu, and D.-S. Zhou, 
J. Phys. A: Math. Theor. {\bf 43} (2010) 315303 
[arXiv:1002.3975 [math-ph]].


\bibitem{Milan} M. Cikovic, Master thesis, April 2011, Bielefeld University.

\bibitem{DLC} R. Delannay and G. Le Ca\"er,
J. Phys. A: Math. Gen {\bf 33} (2000) 2611-2630.

\bibitem{Eugene} E. Strahov, unpublished notes, 2013.

\bibitem{Fischmann}
J. Fischmann, W. Bruzda, B. A. Khoruzhenko, H.-J. Sommers, K. Zyczkowski,
J. Phys. {\bf A45} (2012) 075203 [arXiv:1107.5019 [math-ph]].

\bibitem{Peter1} P.J. Forrester,
J. Phys. {\bf A48} (2015) 215205 [arXiv:1501.05702 [math-ph]].

\bibitem{PeterCoulomb}  P.J. Forrester,
	Nucl. Phys. {\bf B904} (2016) 253-281	[arXiv:1511.02946 [math-ph]].

\bibitem{FI} P.J. Forrester and J.R. Ipsen,
Lin. Alg. Appl.
Vol. {\bf 510} (2016) 259-290 [arXiv:1608.04097 [math-ph]].

\bibitem{FKK} Y.-P. F\"orster, M. Kieburg, and H. K\"osters,
Polynomial Ensembles and P\'olya Frequency Functions,
	arXiv:1710.08794 [math.PR].

\bibitem{FK}
H. Furstenberg and H. Kesten,
Ann. Math. Stat. {\bf 31} (1960) 457-469.

\bibitem{YVF} 
Y.V. Fyodorov, On statistics of bi-orthogonal eigenvectors in real and complex Ginibre ensembles: combining partial Schur decomposition with supersymmetry, 
arXiv:1710.04699 [math-ph].

\bibitem{Ginibre} J. Ginibre, J. Math. Phys. {\bf 6} (1965) 440-449.

\bibitem{GG} F. G\"otze and M. Gorin, 
Commun. Math. Phys. {\bf 281}  (2008) 203-229.
[arXiv: math/0610149].

\bibitem{GT}
F. G\"otze and A. Tikhomirov,
On the Asymptotic Spectrum of Products of Independent Random Matrices,
arXiv:1012.2710v3 [math.PR].

\bibitem{Grad} I.S. Gradshteyn and I.M. Ryzhik, Table of Integrals,
Series, and Products, Academic Press, 6th Edition, San Diego  2000.

\bibitem{Hedenmalm} H. Hedenmalm and A. Wennman,
Planar orthogogonal polynomials and boundary universality in the random normal matrix model,
	arXiv:1710.06493 [math.CV].

\bibitem{Hastings} M. Hastings, 
J. Stat. Phys.  {\bf 103} (2001) 903
[arXiv:cond-mat/9909234].

\bibitem{Ipsen} J.R. Ipsen, 
J. Phys. A: Math. Theor. {\bf 46} (2013) 265201 [arXiv:1301.3343].

\bibitem{IK} J.R. Ipsen and M. Kieburg, Phys. Rev. {\bf E89}  (2014) 032106 [arXiv:1310.4154 [math-ph]].

\bibitem{KK} M. Kieburg and H. K\"osters,
Random Matrices: Theory Appl. {\bf 05} (2016) 1650015 
[arXiv:1601.02586 [math.CA]].

\bibitem{KK2} M. Kieburg and H. K\"osters,
Products of Random Matrices from Polynomial Ensembles,
to appear in Ann. Inst. Henri Poincar\'e Probabilit\'e et Statistique
[arXiv:1601.03724 [math.CA]].

\bibitem{KRV} P. Kopel, S. O'Rourke, and V. Vu,
Random matrix products: Universality and least singular values,
arXiv:1802.03004 [math.PR].


\bibitem{Kostlan}
E. Kostlan, 
Linear Algebra Appl. {\bf 162/164} (1992) 385--388.

\bibitem{LiuZ} D.-Z. Liu and D.-S. Zhou,
J. Stat. Phys. {\bf 140} (2010) 268-288.
[arXiv:0905.4932].

\bibitem{Satya}
S.N. Majumdar, 
Extreme eigenvalues of Wishart matrices: application to entangled bipartite system, 
in: The Oxford Handbook on Random Matrix Theory, G. Akemann, J. Baik, P. Di Francesco (eds.), Oxford University Press, Oxford  2011 [arXiv:1005.4515].

\bibitem{Mehlig} B. Mehlig and J. T. Chalker, 
J. Math. Phys. {\bf 41} (2000)  3233 
[	arXiv:cond-mat/9906279].


\bibitem{Mehta:2004}
M.L.~Mehta, { Random Matrices}, Elsevier, 3rd Edition, Amsterdam 2004.

\bibitem{NIST:2010} F.W.L Olver et al. (eds.), NIST Handbook of Mathematical Functions. Cambridge University Press, Cambridge 2010.

\bibitem{NT}
M.A. Nowak, W. Tarnowski, 
Probing non-orthogonality of eigenvectors in non-Hermitian matrix models: diagrammatic approach, 
arXiv:1801.02526 [math-ph].

\bibitem{Reddy} N.K. Reddy,
Lyapunov exponents and eigenvalues of products of random matrices, arXiv:1606.07704 [math.PR].

\bibitem{Rose}
N. Rosenzweig, 
in: Statistical physics (Brandeis Summer Institute, 1962, Vol. {\bf 3}), New York: W. A. Benjamin (1963)  91-158.

\bibitem{Serfaty} S. Serfaty,
Microscopic description of Log and Coulomb gases,
Lecture notes, Park City mathematics Institute, June 2017, 
arXiv:1709.04089 [math-ph].

\bibitem{TV} T. Tao and V. Vu,
Ann. Probab.
Volume {\bf 43}, Number 2 (2015) 782-874
[arXiv:1206.1893v2 [math.PR]].


\end{thebibliography}
\end{document}